\documentclass[12pt]{article}
\usepackage{a4wide,epsfig}
\usepackage{axodraw4j}
\usepackage{amsmath}
\usepackage{rotating}
\usepackage{slashed}
\usepackage{alltt}
\usepackage{varwidth}
\usepackage{multirow}

\newcommand{\agt}{\rlap{\lower 3.5 pt \hbox{$\mathchar \sim$}} \raise 1pt
 \hbox {$>$}}
\newcommand{\alt}{\rlap{\lower 3.5 pt \hbox{$\mathchar \sim$}} \raise 1pt
 \hbox {$<$}}
\newcommand{\Li}{\mathop{\mathrm{Li}_2}\nolimits}

\catcode`@=11
\newcount\@tempcntc
\def\@citex[#1]#2{\if@filesw\immediate\write\@auxout{\string\citation{#2}}\fi
  \@tempcnta\z@\@tempcntb\m@ne\def\@citea{}\@cite{\@for\@citeb:=#2\do
    {\@ifundefined
       {b@\@citeb}{\@citeo\@tempcntb\m@ne\@citea\def\@citea{,}{\bf
?}\@warning
       {Citation `\@citeb' on page \thepage \space undefined}}%
    {\setbox\z@\hbox{\global\@tempcntc0\csname b@\@citeb\endcsname\relax}%
     \ifnum\@tempcntc=\z@ \@citeo\@tempcntb\m@ne
       \@citea\def\@citea{,}\hbox{\csname b@\@citeb\endcsname}%
     \else
      \advance\@tempcntb\@ne
      \ifnum\@tempcntb=\@tempcntc
      \else\advance\@tempcntb\m@ne\@citeo
      \@tempcnta\@tempcntc\@tempcntb\@tempcntc\fi\fi}}\@citeo}{#1}}
\def\@citeo{\ifnum\@tempcnta>\@tempcntb\else\@citea\def\@citea{,}%
  \ifnum\@tempcnta=\@tempcntb\the\@tempcnta\else
   {\advance\@tempcnta\@ne\ifnum\@tempcnta=\@tempcntb \else
\def\@citea{--}\fi
    \advance\@tempcnta\m@ne\the\@tempcnta\@citea\the\@tempcntb}\fi\fi}
\catcode`@=12

\begin{document}
\allowdisplaybreaks

\title{
\vskip-3cm{\baselineskip14pt
\centerline{\normalsize DESY 19--100\hfill ISSN 0418--9833}
\centerline{\normalsize September 2019\hfill}}
\vskip1.5cm
Dipole subtraction at next-to-leading order in nonrelativistic-QCD
factorization}

\author{Mathias Butenschoen, Bernd A. Kniehl\\
{\normalsize II. Institut f\"ur Theoretische Physik, Universit\"at Hamburg,}\\
{\normalsize Luruper Chaussee 149, 22761 Hamburg, Germany}
}
\date{}

\maketitle

\begin{abstract}
  We describe an implementation of a subtraction scheme in the nonrelativistic-QCD treatment of heavy-quarkonium production at next-to-leading-order in the strong-coupling constant, covering $S$- and $P$-wave bound states. It is based on the dipole subtraction in the massless version by Catani and Seymour and its extension to massive quarks by Phaf and Weinzierl. Important additions include the treatment of heavy-quark bound states, in particular due to the more complicated infrared-divergence structure in the case of $P$-wave states.

\medskip

\noindent
PACS numbers: 12.38.Bx, 12.39.St, 13.85.Ni, 14.40.Pq
\end{abstract}

\newpage


\section{Introduction}

In next-to-leading-order (NLO) perturbative calculations in quantum field theory, the phase space integrations of real corrections generally produce infrared (IR) divergences, which have to be regularized. The standard choice for this is dimensional regularization, where the integrations are done in $D=4-2\epsilon$ space-time dimensions, so that the IR divergences show up as poles in $\epsilon$, ready to be canceled by other contributions. The problem is that the squared matrix elements are, apart from the simplest examples, so complicated that they have to be integrated numerically, in four dimensions. To combine both ingredients, the analytic singularity cancellation in $D$ dimensions and the numerical phase space integration in four dimensions, two basic types of calculational schemes have been devised: slicing schemes and subtraction schemes.

In phase space slicing schemes, the real-correction phase space is split into two parts, with the separation lines enclosing the IR-singular regions at close distances. Since, in the vicinity of the IR divergences, both the squared matrix elements and the phase space factorize into simple expressions, the analytic integration in $D$ dimensions is feasible, while the part outside the enclosed region is free from singularities, ready for numerical integration. Both contributions depend on the specific choice of phase space cut, but the sum of both contributions is independent of it. Most calculations of inclusive heavy-quarkonium production and decay within the factorization formalism \cite{Bodwin:1994jh} of nonrelativistic QCD (NRQCD) \cite{Caswell:1985ui} have been implemented with a two-cutoff phase space slicing scheme as outlined in Ref.~\cite{Harris:2001sx}. In particular, this includes our previous calculations \cite{Klasen:2004tz}. There are, however, two principal disadvantages of the phase space slicing scheme: First, one cannot avoid a residual numerical dependence of the result on the slicing parameters and, second, the numerical integration over the finite real-correction phase space part has to be done to very high precision because there is a strong cancellation between the two phase space parts.

On the other hand, in subtraction schemes, certain simple {\em subtraction terms} with the same divergences as the real corrections are subtracted from the latter, enabling a numerical integration. The subtraction terms are then separately integrated analytically in $D$ dimensions, and the results are added back. To our knowledge, the only NLO calculations of inclusive quarkonium production so far performed in this way are those of Ref.~\cite{Campbell:2007ws} in the color singlet model, based on Catani-Seymour dipole subtraction for massless quarks \cite{Catani:1996vz}. Since only color singlet $S$-wave states were involved, the subtraction terms of Ref.~\cite{Catani:1996vz} were sufficient.

In this paper, we describe an implementation of a subtraction scheme for inclusive quarkonium hadroproduction within NRQCD, which can handle all intermediate $S$- and $P$-wave color singlet and color octet states. In addition to the massless Catani-Seymour scheme \cite{Catani:1996vz}, our implementation is built upon its extension to massive particles by Phaf and Weinzierl~\cite{Phaf:2001gc}. However, we have to take special care of the structures of the amplitudes when projected onto heavy-quark bound states. In particular, new kinds of subtraction terms have to be introduced in the case of $P$-wave state production.

The outline of this paper is as follows: In section~\ref{sec:Structure}, we describe the structure of the appearing amplitudes projected onto the different Fock states and their soft and collinear limits. The divergence cancellation is explained in section~\ref{sec:IRCancellation}. The subtraction scheme used is in detail presented in section~\ref{sec:Subtraction}. Details about the implementation of phase space cuts as well as numerical tests of our extended dipole subtraction approach follow in section~\ref{sec:Implementation}.
Section~\ref{sec:Summary} contains a brief summary.
In Appendix~\ref{sec:AppIntCSPW}, we collect the expressions through order
${\cal O}(\epsilon^0)$ for the integrated Catani-Seymour and Phaf-Weinzierl
dipoles needed in our study, in a form that already includes mass factorization counterterms.

\section{Cross sections and their limits\label{sec:Structure}}

\subsection{Cross sections in NRQCD factorization}

In the framework of QCD and NRQCD factorization, the cross section for the inclusive hadroproduction of heavy quarkonium $H$ is given by
\begin{equation}
d\sigma(AB\to H+X)
=\sum_{a,b,X} \sum_n \int dx_a dx_b\, f_{a/A}(x_a)f_{b/B}(x_b)
\langle{\cal O}^{H}[n]\rangle d\hat{\sigma}(a b\to Q\overline{Q}[n]+X), \label{eq:GeneralFactorization}
\end{equation}
with the partonic cross sections
\begin{eqnarray}
d\hat{\sigma}(a b\to Q\overline{Q}[n]+X) &=&
\frac{1}{N_\mathrm{col}(n) N_\mathrm{pol}(n)}\,\frac{1}{2(p_1+p_2)^2}\, d\mathrm{PS} \nonumber \\
&&\times \frac{F_\mathrm{sym}(X)}{n_\mathrm{col}(a) n_\mathrm{pol}(a) n_\mathrm{col}(b) n_\mathrm{pol}(b)} \| | a b \to Q\overline{Q}[n]+X\rangle \|^2.
\end{eqnarray}
Here, $a$ and $b$ are the colliding QCD partons with four-momenta $p_1$ and $p_2$. $f_{a/A}(x_a)$ is the parton distribution function (PDF) to find parton $a$ with a longitudinal-momentum fraction $x_a$ inside the colliding hadron $A$. $X$ collectively denotes the partons that are produced besides the quarkonium $H$, and $F_\mathrm{sym}(X)$ are its quantum mechanical symmetry factors for identical particles in the final state. $Q$ is bottom for bottomonium production and charm for a charmonium production. $n$ is the $Q\overline{Q}$ Fock state, for our purposes $^3\!S_1$, $^1\!S_0$, $^1\!P_1$, or $^3\!P_J$ in a color singlet or color octet state. The color state is marked by upper indices 1 or 8 in square brackets, like for example in the color octet $^3P_1^{[8]}$ state. $N_\mathrm{col}(n)=1$ if $n$ is a color singlet state and $C_A^2-1=8$ if it is a color octet state, and $N_\mathrm{pol}$ is the $D$-dimensional number of polarization degrees of freedom of state $n$.
We recall that $C_F=4/3$ and $C_A=3$ are color factors of the QCD gauge group
SU(3).
In making the $N_\mathrm{col}$ and $N_\mathrm{pol}$ factors explicit, we follow Ref.~\cite{Petrelli:1997ge}. $\langle{\cal O}^{H}[n]\rangle$ is the corresponding nonperturbative NRQCD long-distance matrix element (LDME). $n_\mathrm{col}(a)$ and $n_\mathrm{pol}(a)$ are the number of colors and the $D$-dimensional number of polarizations of parton $a$. $d\mathrm{PS}$ is the Lorentz-invariant phase space
element. As a convention used throughout this paper, the bra vector is a matrix element, and in squaring the matrix element a summation of the degrees of freedom of all external particles is always understood implicitly. This convention is adopted from Catani, Seymour \cite{Catani:1996vz}, Phaf, and Weinzierl \cite{Phaf:2001gc}, who do, however, include the $n_\mathrm{col}$ factors in the amplitude vectors, albeit not the $n_\mathrm{pol}$ factors. In our choice of normalization, all averaging factors are explicit. Another thing to note is that the summation of external degrees of freedom includes the spin and orbital-angular-momentum quantum numbers $m_s$ and $m_l$ of the $Q\overline{Q}[n]$ state, even if the polarization vectors stand outside the amplitude vectors. Hereby, in the case of a $n={^3P}_J^{[1/8]}$ state, this summation is always restricted to the subspace with definite $J$.

In our study, we are interested in observables where quarkonium $H$ has nonvanishing transverse momentum $p_T$. Therefore, the partonic Born cross sections and their virtual corrections already correspond to $2\to 2$ processes kinematically, namely
\begin{eqnarray}
 g + g&\to&Q\overline{Q}[n]+g, \\
 g + q&\to&Q\overline{Q}[n]+q, \\
 q + g&\to&Q\overline{Q}[n]+q, \\
 q + \overline{q} &\to& Q\overline{Q}[n]+g,
\end{eqnarray}
while, for the real corrections, we are led to consider the $2\to 3$ kinematics processes
\begin{eqnarray}
 g + g&\to&Q\overline{Q}[n]+gg/q\overline{q}, \\
 g + q&\to&Q\overline{Q}[n]+qg, \\
 q + g&\to&Q\overline{Q}[n]+qg, \\
 q + \overline{q} &\to& Q\overline{Q}[n]+gg/q\overline{q}/q'\overline{q'}, \\
 q + q &\to& Q\overline{Q}[n] + q q, \\
 q + q' &\to& Q\overline{Q}[n] + q q',
\end{eqnarray}
where $g$ is a gluon and $q$ a light quark or antiquark (specifically $u$, $d$, $s$, $\overline{u}$, $\overline{d}$, $\overline{s}$ for charmonium and, additionally, $c$ and $\overline{c}$ for bottomonium), $\overline{q}$ its antiparticle, and $q'$ another light quark or antiquark different from $q$ and $\overline{q}$. As already stated above, the four-momenta of the incoming partons are $p_1$ and $p_2$. The four-momenta of the outgoing QCD partons are $p_3$ and, for the real corrections, also $p_4$. The four-momenta of the heavy quark and antiquark that form the $Q\overline{Q}[n]$ state are parameterized by $\frac{p_0}{2}+q$ and $\frac{p_0}{2}-q$, so that $p_0$ is the four-momentum of the $Q\overline{Q}[n]$ state and $2q$ the relative four-momentum of the two constituent heavy quarks. We assume that the mass of the $Q\overline{Q}[n]$ state is twice the heavy-quark mass $m_Q$, $p_0^2=4 m_Q^2$, while we take the other partons to be massless.

The amplitudes $| a b \to Q\overline{Q}[n]+X\rangle$ are evaluated from the usual QCD amplitudes with amputated $Q$ and $\overline{Q}$ spinors $|{\cal A}\rangle$ as
\begin{eqnarray}
 | ^1S_0^{[1/8]} \rangle &=& \mathrm{Tr}\left[{\cal C}_{1/8} \,\Pi_0 \,|{\cal A}\rangle\right]|_{q=0}, \label{Overview.1S08}\\
 | ^3S_1^{[1/8]} \rangle &=& \epsilon_\alpha(m_s) \mathrm{Tr}\left[{\cal C}_{1/8} \,\Pi_1^\alpha\, |{\cal A}\rangle\right]|_{q=0}, \label{Overview.3S11}\\
 | ^1P_1^{[1/8]} \rangle &=& \epsilon_\beta(m_l) \frac{\partial}{\partial q_\beta} \mathrm{Tr}\left[{\cal C}_{1/8} \, \Pi_0 |{\cal A}\rangle\right]|_{q=0}, \\
 | ^3P_J^{[1/8]} \rangle &=& \epsilon_\alpha(m_s) \epsilon_\beta(m_l) \frac{\partial}{\partial q_\beta} \mathrm{Tr}\left[{\cal C}_{1/8} \, \Pi_1^\alpha |{\cal A}\rangle\right]|_{q=0},
\end{eqnarray}
where ${\cal C}_{1}=\frac{1}{\sqrt{2}C_A}$ and ${\cal C}_{8}=\sqrt{2}T_e$ are color projectors with $e$ being the color index of the $c\overline{c}$ color octet state. $\Pi_0$ and $\Pi_1^\alpha$ are the spin projectors \cite{Petrelli:1997ge},
\begin{eqnarray}
 \Pi_0&=&\frac{1}{\sqrt{8m_Q^3}} \left(\frac{\slashed{p}_0}{2}-\slashed{q}-m_Q\right) \gamma_5 \left(\frac{\slashed{p}_0}{2}+\slashed{q}+m_Q\right), \\
 \Pi_1^\alpha
 &=&\frac{1}{\sqrt{8m_Q^3}} \left(\frac{\slashed{p}_0}{2}-\slashed{q}-m_Q\right) \gamma^\alpha \left(\frac{\slashed{p}_0}{2}+\slashed{q}+m_Q\right).
\end{eqnarray}

\subsection{Soft limits\label{sec:softlimits}}

Let us consider a generic Born amplitude,
\begin{equation}
\begin{minipage}{80pt}
\begin{picture}(80,50)(0,0)
\GOval(13,25)(10,10)(0){0.9}
\DashLine(73,25)(23,25){2}
\Text(48,33)[b]{$p_i$}
\Text(48,29)[]{$\rightarrow$}
\end{picture}
\end{minipage}
= |\mathrm{Born}\rangle.
\end{equation}
Then, the expression for the scattering amplitude with an additional gluon $j$ with momentum $p_j$ attached to the line of an outgoing parton $i$ is in the limit of $p_j$ being soft given by the eikonal approximation,
\begin{equation}
\left.
\begin{minipage}{80pt}
\begin{picture}(80,50)(0,-10)
\GOval(13,25)(10,10)(0){0.9}
\DashLine(73,25)(23,25){2}
\Text(34,33)[b]{\scriptsize $p_i+p_j$}
\Text(34,29)[]{$\rightarrow$}
\Text(59,33)[b]{\scriptsize $p_i$}
\Text(59,29)[]{$\rightarrow$}
\Gluon(43,25)(73,5){2}{8}
\Text(70,-1)[t]{\scriptsize $p_j$}
\Text(70,0)[]{$\rightarrow$}
\Text(53,13)[rt]{\scriptsize$c$}
\Text(27.5,22)[t]{\scriptsize $a$}
\Text(70,22)[t]{\scriptsize $b$}
\end{picture}
\end{minipage}
\right|_{p_j\;\mathrm{soft}} = g_s \frac{p_i\cdot\epsilon^\ast(p_j)}{p_i\cdot p_j} \mathbf{T}_i |\mathrm{Born}\rangle.
\label{eq:EikonalApprox}
\end{equation}
Here, $g_s$ is the QCD gauge coupling, and $\mathbf{T}_i$ acts on $|A\rangle$ by inserting at the appropriate place $T_c$ if parton $i$ is an outgoing quark or incoming antiquark, $-T_c$ if parton $i$ is an incoming antiquark or outgoing quark, and $if_{abc}$ if parton $i$ is a gluon.

Let us now consider a generic real-correction amplitude in the limit where a certain gluon $j$ with momentum $p_j$ is soft,
\begin{equation}
\left.
\begin{minipage}{104pt}
\begin{picture}(104,62) (-10,-16)
\DashLine(-11,35)(25,35){2}
\DashLine(-11,10)(25,10){2}
\ArrowLine(50,35)(86,35)
\ArrowLine(86,22.5)(55.1777,22.5)
\DashLine(50,10)(86,10){2}
\Gluon(40,5)(86,-2.5){2}{9}
\Text(82,-7)[]{\small $\rightarrow$}
\Text(82,-8)[t]{\scriptsize $p_j$}
\GOval(37.5,22.5)(17.6777,17.6777)(0){0.9}
\Text(37.5,22.5)[]{Real}
\Text(88,35)[l]{\scriptsize $c$}
\Text(88,22.5)[l]{\scriptsize $\overline{c}$}
\end{picture}
\end{minipage}
\right|_{p_j\;\mathrm{soft}} = |p_j\;\mathrm{soft}\rangle.
\end{equation}
Since this implies the sum of all those diagrams where gluon $j$ is in turn attached to all the other external-particle lines of the corresponding Born diagrams, application of Eq.~(\ref{eq:EikonalApprox}) yields
\begin{equation}
 |p_j\;\mathrm{soft}\rangle = g_s\Bigg(\frac{\left(\frac{p_0}{2}+q\right)\cdot\epsilon^\ast(p_j)}{\left(\frac{p_0}{2}+q\right)\cdot p_j} \mathbf{T}_c + \frac{\left(\frac{p_0}{2}-q\right)\cdot\epsilon^\ast(p_j)}{\left(\frac{p_0}{2}-q\right)\cdot p_j} \mathbf{T}_{\overline{c}} + \sum_{\substack{i=1 \\ i\neq j}}^4 \frac{p_i\cdot\epsilon^\ast(p_j)}{p_i\cdot p_j} \mathbf{T}_i \Bigg) \big|\mathrm{Born}\big\rangle.
\end{equation}
For definiteness, we evaluate these soft limits using the axial gauge, $p_0\cdot\epsilon(p_j)=0$, so that the gluon polarization sum takes the form
\begin{equation}
 \Pi_{\alpha\beta}(p_j) \equiv \sum_{\mathrm{pol}} \epsilon_\alpha(p_j) \epsilon_\beta^\ast(p_j) = -g_{\alpha\beta} + \frac{p_{0\alpha} p_{j\beta} + p_{j\alpha}p_{0\beta}}{p_0\cdot p_j} - \frac{p_0^2 p_{j\alpha} p_{j\beta}}{(p_0\cdot p_j)^2}. \label{eq:axialgaugepolsum}
\end{equation}
Applying the projectors and squaring the matrix elements then yields
\begin{eqnarray}
\| |^1\!S_0^{[1/8]},p_j\;\mathrm{soft}\rangle \|^2 &=& S_1 (^1\!S_0^{[1/8]}; p_j), \\
\| |^3\!S_1^{[1/8]},p_j\;\mathrm{soft}\rangle \|^2 &=& S_1 (^3\!S_1^{[1/8]}; p_j), \\
\| |^1\!P_1^{[1/8]},p_j\;\mathrm{soft}\rangle \|^2
 &=& S_1 (^1\!P_1^{[1/8]}; p_j) + S_2(^1\!P_1^{[1/8]}, ^1\!S_0^{[1/8]}; p_j) + S_3(^1\!S_0^{[1/8]}; p_j), \\
\| |^3\!P_J^{[1/8]},p_j\;\mathrm{soft}\rangle \|^2
 &=& S_1 (^3\!P_J^{[1/8]}; p_j) + S_2(^3\!P_J^{[1/8]}, ^3\!S_1^{[1/8]}; p_j) + S_3(^3\!S_1^{[1/8]}; p_j),
\end{eqnarray}
with
\begin{eqnarray}
 S_1(n; p_j) &=& g_s^2\sum_{\substack{i,k=1 \\ i,k\neq j}}^4 \frac{\Pi^{\alpha\beta}(p_j) p_{i\alpha} p_{k\beta} }{p_i\cdot p_j \; p_k\cdot p_j}
\langle \mbox{$n$, Born} | \mathbf{T}_i \mathbf{T}_k | \mbox{$n$, Born} \rangle, \label{eq:s1term}\\
S_2(n,m; p_j) &=& 4 g_s^2 \sum_{\substack{i=1 \\ i\neq j}}^4 \frac{\Pi^{\alpha\beta}(p_j) p_{i\alpha} } {p_i\cdot p_j \; p_0\cdot p_j} \epsilon_\beta(m_l) \langle n,\mathrm{Born} | \mathbf{T}_i (\mathbf{T}_c-\mathbf{T}_{\overline{c}}) | m,\mathrm{Born}\rangle,
\\
 S_3(m; p_j) &=& 4g_s^2 \frac{ \Pi^{\alpha\beta}(p_j) } {(p_0\cdot p_j)^2}
\epsilon^\ast_\alpha(m_l)\epsilon_\beta(m_l) \langle m,\mathrm{Born} |  (\mathbf{T}_c-\mathbf{T}_{\overline{c}}) (\mathbf{T}_c-\mathbf{T}_{\overline{c}}) | m,\mathrm{Born}\rangle.
\end{eqnarray}
Defining $\mathbf{T}_0\equiv-\mathbf{T}_1-\mathbf{T}_2-\mathbf{T}_3=\mathbf{T}_c+\mathbf{T}_{\overline{c}}$, we can write
\begin{eqnarray}
S_1(n; p_j)
&=&  g_s^2 \sum_{\substack{k=1 \\ k\neq j}}^4  \langle n,\mathrm{Born}|\left( -\sum_{\substack{i=1 \\ i\neq j}}^4 \frac{p_i\cdot p_k}{p_i\cdot p_j \;p_k\cdot p_j} \mathbf{T}_i \mathbf{T}_k -\frac{2 p_0\cdot p_k}{p_k\cdot p_j\; p_0\cdot p_j} \right.  \mathbf{T}_0 \mathbf{T}_k \nonumber \\ && \left. + \frac{p_0^2}{(p_0\cdot p_j)^2} \mathbf{T}_0 \mathbf{T}_k\right) | n, \mathrm{Born} \rangle \nonumber \\
&=& g_s^2 \sum_{\substack{i=0 \\ i\neq j}}^4 \sum_{\substack{k=0 \\ k\neq i,j}}^4 \left(-\frac{p_i\cdot p_k}{p_i \cdot p_j\; p_k\cdot p_j} + \frac{p_i^2}{(p_i\cdot p_j)^2} \right) \langle n,\mathrm{Born} | \mathbf{T}_i \mathbf{T}_k | n, \mathrm{Born} \rangle \nonumber \\
&=& -g_s^2 \sum_{\substack{i=0 \\ i\neq j}}^4 \sum_{\substack{k=0 \\ k\neq i,j}}^4 \left(\frac{2p_i\cdot p_k}{p_i \cdot p_j\; (p_i+p_k)\cdot p_j} - \frac{p_i^2}{(p_i\cdot p_j)^2} \right) \langle n, \mathrm{Born} | \mathbf{T}_i \mathbf{T}_k | n, \mathrm{Born} \rangle,\quad \label{eq:soft1singular}
\\
S_2(n,m; p_j) &=& 4g_s^2\sum_{\substack{i=1 \\ i\neq j}}^4 \left( \frac{-p_i^\beta}{p_i\cdot p_j\; p_0\cdot p_j} + \frac{p_0\cdot p_i\; p_j^\beta}{p_i\cdot p_j (p_0\cdot p_j)^2} -\frac{p_0^2 p_j^\beta}{(p_0\cdot p_j)^3} \right) \nonumber \\
&&\times \epsilon_\beta(m_l) \langle n,\mathrm{Born} | \mathbf{T}_i (\mathbf{T}_c-\mathbf{T}_{\overline{c}}) | m,\mathrm{Born}\rangle, \label{eq:s2term}
\\
S_3(m; p_j) &=& 4 g_s^2 \left( - \frac{g^{\alpha\beta}}{(p_0\cdot p_j)^2} - \frac{p_0^2 p_j^\alpha p_j^\beta}{(p_0\cdot p_j)^4} \right) \nonumber
\\
&&\times \epsilon^\ast_\alpha(m_l)\epsilon_\beta(m_l) \langle m,\mathrm{Born} |  (\mathbf{T}_c-\mathbf{T}_{\overline{c}}) (\mathbf{T}_c-\mathbf{T}_{\overline{c}}) | m,\mathrm{Born}\rangle \label{eq:s3term}.
\end{eqnarray}
The $S_2$ and $S_3$ terms, which only appear in squared amplitudes of $P$-wave states, are specific for our study.

\subsection{Collinear limits\label{sec:collinearlimits}}

Let us first consider the limit where an incoming parton with momentum $p_i$ is collinear to the outgoing parton with momentum $p_j$. In this limit the divergent contributions stem from the diagrams with $i\to (ij)+j$ splitting,
\begin{equation}
\begin{minipage}{123pt}
\begin{picture}(123,70)(-11,-6)
\DashLine(-11,47)(25,47){2}
\DashLine(-11,22)(25,22){2}
\DashCArc(25,22)(18,180,270){2}
\DashLine(25,4)(86,4){2}
\ArrowLine(50,47)(86,47)
\ArrowLine(86,34.5)(55.1777,34.5)
\DashLine(50,22)(86,22){2}
\GOval(37.5,34.5)(17.6777,17.6777)(0){0.9}
\Text(37.5,34.5)[]{Born}
\Text(88,4)[l]{$\rightarrow$\footnotesize$\,p_j$}
\Text(-2,26.5)[b]{\scriptsize $p_i$}
\Text(-2,24)[]{$\rightarrow$}
\Text(15,26.5)[b]{\scriptsize $p_{ij}$}
\Text(15,24)[]{$\rightarrow$}
\end{picture}
\end{minipage}
\;.
\end{equation}
If we define $p_\perp$ to be the transverse momentum of $p_{(ij)}$ in the rest frame of the incoming partons, then we can define the fraction $x$ of the longitudinal momentum of $i$ taken away by $(ij)$ as
\begin{eqnarray}
 p_{(ij)} &=& x p_i + {\cal O}(p_\perp), \\
 p_j &=& (1-x) p_i + {\cal O}(p_\perp), \\
 2 p_i\cdot p_j &=& \frac{-p_\perp^2}{1-x} + {\cal O} (p_\perp^3).
\end{eqnarray}
Then, the squared matrix element factorizes in the collinear limit as
\begin{eqnarray}
 \| |p_j \;\mathrm{ini.\;coll.}\; p_i \rangle \|^2 &=& \frac{n_\mathrm{col}(i)
 }{n_\mathrm{col}((ij)) n_\mathrm{pol}((ij)) } \, \frac{g_s^2}{x(p_i\cdot p_j)} \,
 \langle \mathrm{Born} | \hat{P}_{i,(ij)}(x,p_\perp) | \mathrm{Born} \rangle \nonumber \\
 &&\times
 \begin{cases}
  \delta_{ss'} & \mbox{if $i$ is a quark or antiquark} \\
  \epsilon_\mu^\ast(p_i) \epsilon_\nu(p_i) & \mbox {if $i$ is a gluon}
 \end{cases},
\end{eqnarray}
where the indices $s$ and $s'$ or $\mu$ and $\nu$ are the spin or polarization indices of particle $i$, and $\hat{P}_{i,(ij)}(x)$ are the spin-dependent $D$-dimensional Dokshitzer-Gribov-Lipatov-Altarelli-Parisi (DGLAP) \cite{Gribov:1972ri} splitting functions, which, up to order ${\cal O}(\epsilon)$, are given by
\begin{eqnarray}
 \hat{P}_{qq}(x,p_\perp) &=& \delta_{ss'}C_F \left(\frac{1+x^2}{1-x}-\epsilon(1-x)\right),
 \\
 \hat{P}_{qg}(x,p_\perp) &=& \delta_{ss'}C_F \left(\frac{1+(1-x)^2}{x}-\epsilon x\right),
 \\
 \hat{P}_{gq}(x,p_\perp) &=& \frac{1}{2} \left( -g^{\mu\nu} + 4 x(1-x) \frac{p_\perp^\mu p_\perp^\nu}{p_\perp^2} \right),
 \\
 \hat{P}_{gg}(x,p_\perp) &=& 2 C_A \left( -g^{\mu\nu}\left( \frac{x}{1-x} + \frac{1-x}{x}\right) - 2(1-\epsilon)x(1-x)\frac{p_\perp^\mu p_\perp^\nu}{p_\perp^2} \right).
\end{eqnarray}
Here, $(ij)$ and $j$ are labeled $q$ if the corresponding particle is a quark or antiquark and $g$ if it is a gluon. We insert unity noticing that $\mathbf{T}_{(ij)}=-\sum_{\substack{k=0\\k\neq i,j}}^4 \mathbf{T}_k$ and so obtain
\begin{eqnarray}
\| |p_j \;\mathrm{ini.\;coll.}\; p_i \rangle \|^2 &=& \frac{n_\mathrm{col}(i)
 }{n_\mathrm{col}((ij)) n_\mathrm{pol}((ij)) } \, \frac{-g_s^2}{x(p_i\cdot p_j)}
 \sum_{\substack{k=0\\k\neq i,j}}^4 \langle \mathrm{Born} | \hat{P}_{i,(ij)}(x,p_\perp) \frac{\mathbf{T}_{(ij)} \mathbf{T}_k}{\mathbf{T}_{(ij)}^2} | \mathrm{Born} \rangle\nonumber \\
 &&\times
 \begin{cases}
  \delta_{ss'} & \mbox{if $i$ is a quark or antiquark} \\
  \epsilon_\mu^\ast(p_i) \epsilon_\nu(p_i) & \mbox {if $i$ is a gluon}
 \end{cases}, \label{eq:IniColl}
\end{eqnarray}
where we note that $\mathbf{T}_{(ij)}^2=C_F$ if $(ij)$ is a quark or antiquark, and $C_A$ if $(ij)$ is a gluon.
 
If the two outgoing partons 3 and 4 are collinear, the dominant contributions stem from diagrams where there is a $(34)\to 3 + 4$ splitting,
\begin{equation}
\begin{minipage}{123pt}
\begin{picture}(123,70)(-11,-6)
\DashLine(-11,47)(25,47){2}
\DashLine(-11,22)(25,22){2}
\ArrowLine(50,47)(86,47)
\ArrowLine(86,34.5)(55.1777,34.5)
\DashLine(50,22)(86,22){2}
\DashLine(68,22)(86,4){2}
\GOval(37.5,34.5)(17.6777,17.6777)(0){0.9}
\Text(37.5,34.5)[]{Born}
\Text(88,4)[l]{$\rightarrow$\footnotesize$\,p_4$}
\Text(88,22)[l]{$\rightarrow$\footnotesize$\,p_3$}
\Text(61,13)[t]{\scriptsize $p_{34}$}
\Text(61,16)[]{$\rightarrow$}
\end{picture}
\end{minipage}
\;.
\end{equation}
If we define $p_\perp$ to be the transverse momentum of $p_{3}$ in the rest frame of the incoming partons, then we can define the fraction $z$ of the longitudinal momentum of $(34)$ taken away by $3$ as
\begin{eqnarray}
 p_3 &=& z p_{(34)} + {\cal O}(p_\perp), \\
 p_4 &=& (1-z) p_{(34)} + {\cal O}(p_\perp), \\
 2 p_3\cdot p_4 &=& \frac{-p_\perp^2}{z(1-z)} + {\cal O} (p_\perp^3).
\end{eqnarray}
The squared matrix element then factorizes as
\begin{eqnarray}
 \| |p_3 \;\mathrm{final\;coll.}\; p_4 \rangle \|^2 &=& \frac{g_s^2}{p_3\cdot p_4}
 \langle \mathrm{Born} | \hat{P}_{(34),3}(z,p_\perp) | \mathrm{Born} \rangle
 \nonumber \\
 &=& - \frac{g_s^2}{p_3\cdot p_4} \sum_{\substack{k=0 \\ k\neq 3,4}}^4 \langle \mathrm{Born} | \hat{P}_{(34),3}(z,p_\perp) \frac{\mathbf{T}_{(34)}\mathbf{T}_k}{\mathbf{T}_{(34)}^2} | \mathrm{Born} \rangle
 , \label{eq:FinColl}
\end{eqnarray}
where the indices $s$ and $s'$ or $\mu$ and $\nu$ within $\hat{P}_{(34),3}$ are the open spin or Lorentz indices of particle $(34)$ in the Born amplitude.

\section{Cancellation of IR divergences\label{sec:IRCancellation}}

\begin{figure}
\begin{center}
\includegraphics[width=\textwidth]{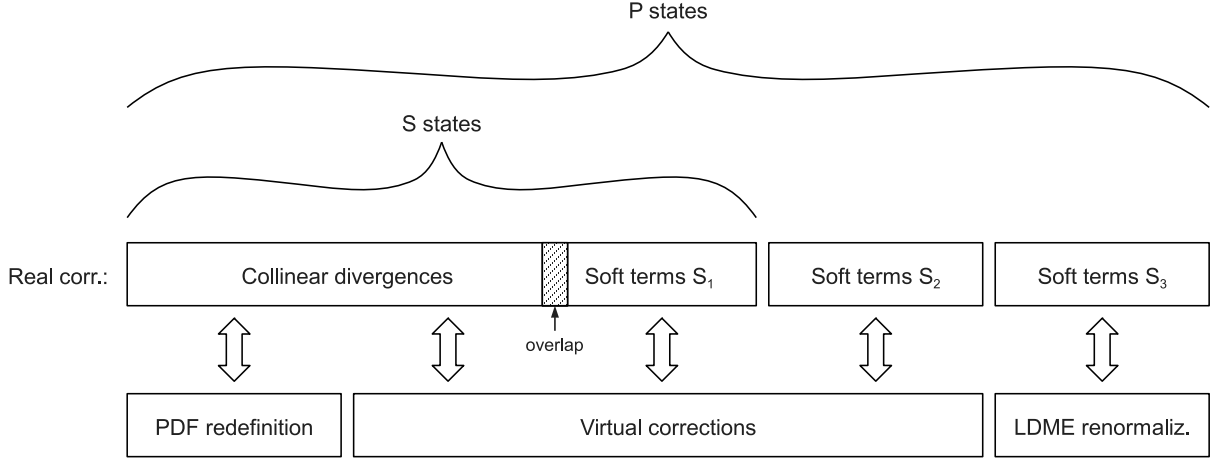}
\end{center}
\caption{\label{SingStructure}Overview of the IR-singularity structure and its cancellations.}
\end{figure}

The IR divergences associated with the soft and collinear limits discussed in sections~\ref{sec:softlimits} and~\ref{sec:collinearlimits} are to a large extent canceled by contributions of the virtual corrections, as shown in Fig.~\ref{SingStructure}. A part of the initial-state collinear divergences is, however, absorbed in the PDFs, while the $S_3$ contributions to the soft divergences are canceled by LDME renormalization contributions. These two additional ingredients are described in this section.

\subsection{PDF redefinition and PDF evolution\label{sec:MFC}}

As for the initial-state collinear divergences of an $i\to (ij)+j$ splitting, a part of it is absorbed by an $\overline{\mathrm{MS}}$ redefinition of the PDF $f_{(ij)/A}(y)$, which then becomes dependent on the factorization scale $\mu_f$,
\begin{equation}
 f_{(ij)/A}(y,\mu_f) \equiv f_{(ij)/A}(y) - \frac{g_s^2}{8\pi^2}  \left( \frac{4\pi\mu_r^2}{\mu_f^2} e^{-\gamma_E} \right)^\epsilon \frac{1}{\epsilon} \int_y^1 \frac{dx}{x} P_{i,(ij)}^+(x) f_{i/A}\left(\frac{y}{x}\right),
 \label{eq:FacDepPDF}
\end{equation}
where $\mu_r$ is the renormalization scale and $P_{i,(ij)}^+(x)$ is one of the regularized DGLAP splitting functions,
\begin{eqnarray}
P_{qq}^+(x) &=& C_F \left( \frac{1+x^2}{(1-x)_+} + \frac{3}{2}\delta(1-x) \right), \\
P_{qg}^+(x) &=& C_F \frac{1+(1-x)^2}{x}, \\
P_{gq}^+(x) &=& \frac{1}{2}\left( x^2+(1-x)^2 \right), \\
P_{gg}^+(x) &=& 2 C_A \left( \frac{x}{(1-x)_+} + \frac{1-x}{x} + x(1-x) \right) + \left( \frac{11}{6}C_A - \frac{n_f}{3}\right) \delta(1-x),
\end{eqnarray}
with $n_f$ being the number of active quark flavors, for us 3 for charmonium production and 4 for bottomonium production. Next, we solve Eq.~(\ref{eq:FacDepPDF}) for $f_{(ij)/A}(y)$. Using these $f_{(ij)/A}(y)$ functions in the general formula (\ref{eq:GeneralFactorization}), a {\em mass factorization counterterm},
\begin{eqnarray}
 &&d\hat\sigma_\mathrm{MFC}(a+b+Q\overline{Q}[n]+X) = \left[ \sum_{(ij)} \int dx P_{a,(ij)}^+(x) d\hat\sigma_\mathrm{Born}((ij)+b\to Q\overline{Q}[n]+X) \right. \nonumber \\
 && \quad+ \left. \sum_{(ij)} \int dx P_{b,(ij)}^+(x) d\hat\sigma_\mathrm{Born}(a+(ij)\to Q\overline{Q}[n]+X)\right] \frac{g_s^2}{8\pi^2} \left( \frac{4\pi\mu_r^2}{\mu_f^2}e^{-\gamma_E}\right)^\epsilon \frac{1}{\epsilon},\label{eq:MFC}
\end{eqnarray}
arises, where parton $(ij)$ carries the fraction $x$ of the splitting parton's momentum. In Fig.~\ref{SingStructure}, this contribution is indicated as the box labeled {\em PDF redefinition}. The DGLAP equations governing the evolution of the scale-dependent PDFs follow seamlessly after differentiating Eq.~(\ref{eq:FacDepPDF}) with respect to $\mu_f$.

\subsection{LDME renormalization and LDME evolution}

In a similar fashion, we have to treat the contributions from the LDME renormalization. A NLO calculation of the $^3S_1^{[8]}$ LDME using NRQCD Feynman rules and an expansion in $\frac{1}{m_Q}$ after the loop integration yields that it is related to the $^3P_J^{[1]}$ and $^3P_J^{[8]}$ LDMEs via
\begin{eqnarray}
\langle {\cal O}^{H}[^3S_1^{[8]}]\rangle_{\mathrm{NLO}} &=& \langle {\cal O}^{H}[^3S_1^{[8]}]\rangle_{\mathrm{LO}}
+ \frac{g_s^2}{3\pi^2 m_Q^2} \left( \frac{1}{\epsilon_{\mbox{\tiny UV}}}-\frac{1}{\epsilon_{\mbox{\tiny IR}}} \right) \nonumber \\
&&\times \sum_J
\left[ \frac{C_F}{2C_A} \langle {\cal O}^{H}[^3P_J^{[1]}]\rangle
 + \frac{C_A^2-4}{4C_A} \langle {\cal O}^{H}[^3P_J^{[8]}]\rangle \right]. \label{eq:NLOLDME1}
\end{eqnarray}
This bare operator is both ultraviolet and IR divergent. We remove the ultraviolet singularity by introducing an $\overline{\mathrm{MS}}$-renormalized LDME,
\begin{eqnarray}
\langle {\cal O}^{H}[^3S_1^{[8]}]\rangle^{(\mu_\Lambda)} &\equiv& \langle {\cal O}^{H}[^3S_1^{[8]}]\rangle_{\mathrm{NLO}} - \frac{g_s^2}{3\pi^2 m_Q^2} \left( \frac{4\pi\mu_r^2}{\mu_\Lambda^2} e^{-\gamma_E} \right)^\epsilon \frac{1}{\epsilon_{\mbox{\tiny UV}}} 
\nonumber \\
&&\times \sum_J
\left[ \frac{C_F}{2C_A} \langle {\cal O}^{H}[^3P_J^{[1]}]\rangle
 + \frac{C_A^2-4}{4C_A} \langle {\cal O}^{H}[^3P_J^{[8]}]\rangle \right], \label{eq:NLOLDME2}
\end{eqnarray}
which depends on the NRQCD factorization scale $\mu_\Lambda$.
Solving Eqs.~(\ref{eq:NLOLDME1}) and (\ref{eq:NLOLDME2}) for $\langle {\cal O}^{H}[^3S_1^{[8]}]\rangle_{\mathrm{LO}}$ and using this in the general formula (\ref{eq:GeneralFactorization}), we obtain the contribution
\begin{eqnarray}
 d\sigma_{^3S_1^{[8]}\, \mathrm{op.ren.}} &=& \sum_{\substack{a,b,X\\2\to 2}}\int dx_a dx_b f_{a/A}(x_a) f_{b/B}(x_b) d\hat{\sigma}(ab\to Q\overline{Q}[^3S_1^{[8]}]+X)   \nonumber
 \\
 &&\times \frac{g_s^2}{3\pi^2 m_Q^2} \left( \frac{4\pi\mu_r^2}{\mu_\Lambda^2} e^{-\gamma_E} \right)^\epsilon \frac{1}{\epsilon}\sum_J
\left[ \frac{C_F}{2C_A} \langle {\cal O}^{H}[^3P_J^{[1]}]\rangle
 + \frac{C_A^2-4}{4C_A} \langle {\cal O}^{H}[^3P_J^{[8]}]\rangle \right].\qquad \label{eq:opren3s18}
\end{eqnarray}
Similarly, we obtain
\begin{eqnarray}
 d\sigma_{^3S_1^{[1]}\, \mathrm{op.ren.}} &=& \sum_{\substack{a,b,X\\2\to 2}} \int dx_a dx_b f_{a/A}(x_a) f_{b/B}(x_b) d\hat{\sigma}(ab\to Q\overline{Q}[^3S_1^{[1]}]+X)   \nonumber
 \\
 &&\times \frac{g_s^2}{3\pi^2 m_Q^2} \left( \frac{4\pi\mu_r^2}{\mu_\Lambda^2} e^{-\gamma_E} \right)^\epsilon \frac{1}{\epsilon}\sum_J
\langle {\cal O}^{H}[^3P_J^{[8]}]\rangle, \label{eq:opren3s11}
\\
d\sigma_{^1S_0^{[8]}\, \mathrm{op.ren.}} &=& \sum_{\substack{a,b,X\\2\to 2}} \int dx_a dx_b f_{a/A}(x_a) f_{b/B}(x_b) d\hat{\sigma}(ab\to Q\overline{Q}[^1S_0^{[8]}]+X)   \nonumber
 \\
 &&\times \frac{g_s^2}{3\pi^2 m_Q^2} \left( \frac{4\pi\mu_r^2}{\mu_\Lambda^2} e^{-\gamma_E} \right)^\epsilon \frac{1}{\epsilon}
\left[ \frac{C_F}{2C_A} \langle {\cal O}^{H}[^1P_1^{[1]}]\rangle
 + \frac{C_A^2-4}{4C_A} \langle {\cal O}^{H}[^1P_1^{[8]}]\rangle \right], \label{eq:opren1s08}
\\
d\sigma_{^1S_0^{[1]}\, \mathrm{op.ren.}} &=&  \sum_{\substack{a,b,X\\2\to 2}}\int dx_a dx_b f_{a/A}(x_a) f_{b/B}(x_b) d\hat{\sigma}(ab\to Q\overline{Q}[^1S_0^{[1]}]+X)  \nonumber
 \\
 &&\times \frac{g_s^2}{3\pi^2 m_Q^2} \left( \frac{4\pi\mu_r^2}{\mu_\Lambda^2} e^{-\gamma_E} \right)^\epsilon \frac{1}{\epsilon}\langle {\cal O}^{H}[^1P_1^{[8]}]\rangle. \label{eq:opren1s01}
\end{eqnarray}
These contributions cancel the $S_3$ contributions of the soft divergences and are labeled {\em LDME renormalization} in Fig.~\ref{SingStructure}. However, we transform them further to cast them into a form that will be more useful for our purposes. Noticing that
\begin{eqnarray}
\langle ^{2S+1}L_J^{[8]} | (\mathbf{T}_c-\mathbf{T}_{\overline{c}}) (\mathbf{T}_c-\mathbf{T}_{\overline{c}}) | ^{2S+1}L_J^{[8]} \rangle &=& \frac{C_A^2-4}{C_A} \| | ^{2S+1}L_J^{[8]} \rangle \|^2 + 8 C_A C_F \| | ^{2S+1}L_J^{[1]} \rangle \|^2, \\
\langle ^{2S+1}L_J^{[1]} | (\mathbf{T}_c-\mathbf{T}_{\overline{c}}) (\mathbf{T}_c-\mathbf{T}_{\overline{c}}) | ^{2S+1}L_J^{[1]} \rangle &=& \frac{1}{C_A^2} \| | ^{2S+1}L_J^{[8]} \rangle \|^2, 
\end{eqnarray}
we can rewrite Eqs.~(\ref{eq:opren3s18}) and (\ref{eq:opren3s11}) as
\begin{eqnarray}
&&d\sigma_{^3S_1^{[1]}+^3S_1^{[8]}\, \mathrm{op.ren.}} = \frac{g_s^2}{12\pi^2 m_Q^2} \left( \frac{4\pi\mu_r^2}{\mu_\Lambda^2} e^{-\gamma_E} \right)^\epsilon \frac{1}{\epsilon} \sum_{\substack{a,b,X\\2\to 2}} \int dx_a dx_b\, f_{a/A}(x_a)f_{b/B}(x_b) \, \frac{1}{2(p_1+p_2)^2}\,   \nonumber \\
\lefteqn{ \qquad \times  \frac{F_\mathrm{sym}(X)}{n_\mathrm{col}(a) n_\mathrm{pol}(a) n_\mathrm{col}(b) n_\mathrm{pol}(b)} \,d\mathrm{PS}_2 \sum_J \left[ \frac{\langle ^3S_1^{[1]}, \mathrm{Born} | (\mathbf{T}_c-\mathbf{T}_{\overline{c}}) (\mathbf{T}_c-\mathbf{T}_{\overline{c}}) | ^3S_1^{[1]}, \mathrm{Born} \rangle}{N_\mathrm{col}(^3S_1^{[1]}) N_\mathrm{pol}(^3S_1^{[1]})}   \right.} \nonumber \\
\lefteqn{\left.\qquad \times \langle{\cal O}^{H}[^3P_J^{[1]}]\rangle + \frac{\langle ^3S_1^{[8]}, \mathrm{Born} | (\mathbf{T}_c-\mathbf{T}_{\overline{c}}) (\mathbf{T}_c-\mathbf{T}_{\overline{c}}) | ^3S_1^{[8]} , \mathrm{Born}\rangle}{N_\mathrm{col}(^3S_1^{[8]}) N_\mathrm{pol}(^3S_1^{[8]})}  \langle {\cal O}^{H}[^3P_J^{[8]}]\rangle \right]. }
\end{eqnarray}
Recalling our convention regarding the summation of the polarization degrees of freedom, we observe that $N_\mathrm{pol}(^3S_1)=-\epsilon_\mu^\ast(m_s)\epsilon^\mu(m_s)$ and $N_\mathrm{pol}(^3P_J)=\epsilon_\mu^\ast(m_l)\epsilon_\nu^\ast(m_s)\epsilon^\mu(m_l)\epsilon^\nu(m_s)$, so that we can write
\begin{eqnarray}
d\sigma_{^3S_1^{[1]}+^3S_1^{[8]}\, \mathrm{op.ren.}}
&=&\sum_{\substack{a,b,X\\2\to 2}}\sum_{c=1,8}\sum_J \int dx_a dx_b\, f_{a/A}(x_a)f_{b/B}(x_b)
\frac{\langle{\cal O}^{H}[^3P_J^{[c]}]\rangle}{N_\mathrm{col}(^3P_J^{[c]}) N_\mathrm{pol}(^3P_J^{[c]})} \nonumber \\
&&\times \frac{1}{2(p_1+p_2)^2}\, d\mathrm{PS}_2
\frac{F_\mathrm{sym}(X)}{n_\mathrm{col}(a) n_\mathrm{pol}(a) n_\mathrm{col}(b) n_\mathrm{pol}(b)} \| | ^3P_J^{[c]},\,\mathrm{op.ren.}\rangle \|^2,\quad
\label{eq:OpRenTransformed3PJOne}
\end{eqnarray}
with
\begin{eqnarray}
 \| | ^3P_J^{[c]},\mathrm{op.ren.}\rangle \|^2 &=& \frac{g_s^2}{12\pi^2 m_Q^2} \left( \frac{4\pi\mu_r^2}{\mu_\Lambda^2} e^{-\gamma_E} \right)^\epsilon g^{\alpha\beta}\left(-\frac{1}{\epsilon}\right) \nonumber \\ &&\times \epsilon_\alpha^\ast(m_l)\epsilon_\beta(m_l) \langle ^3S_1^{[c]}, \mathrm{Born} | (\mathbf{T}_c-\mathbf{T}_{\overline{c}}) (\mathbf{T}_c-\mathbf{T}_{\overline{c}}) | ^3S_1^{[c]}, \mathrm{Born} \rangle.
 \label{eq:OpRenTransformed3PJTwo}
\end{eqnarray}
From the terms in Eqs.~(\ref{eq:opren1s08}) and (\ref{eq:opren1s01}), we obtain a corresponding expression,
\begin{eqnarray}
d\sigma_{^1S_0^{[1]}+^1S_0^{[8]}\, \mathrm{op.ren.}}
&=&\sum_{\substack{a,b,X\\2\to 2}}\sum_{c=1,8} \int dx_a dx_b\, f_{a/A}(x_a)f_{b/B}(x_b)
\frac{\langle{\cal O}^{H}[^1P_1^{[c]}]\rangle}{N_\mathrm{col}(^1P_1^{[c]}) N_\mathrm{pol}(^1P_1^{[c]})} \nonumber \\
&&\times \frac{1}{2(p_1+p_2)^2}\, d\mathrm{PS}_2
\frac{F_\mathrm{sym}(X)}{n_\mathrm{col}(a) n_\mathrm{pol}(a) n_\mathrm{col}(b) n_\mathrm{pol}(b)} \| | ^1P_1^{[c]},\,\mathrm{op.ren.}\rangle \|^2,\quad
\label{eq:OpRenTransformed1P1One}
\end{eqnarray}
with
\begin{eqnarray}
 \| | ^1P_1^{[c]},\mathrm{op.ren.}\rangle \|^2 &=& \frac{g_s^2}{12\pi^2 m_Q^2} \left( \frac{4\pi\mu_r^2}{\mu_\Lambda^2} e^{-\gamma_E} \right)^\epsilon g^{\alpha\beta} \left(-\frac{1}{\epsilon}\right) \nonumber \\ &&\times \epsilon_\alpha^\ast(m_l)\epsilon_\beta(m_l) \langle ^1S_0^{[c]}, \mathrm{Born} | (\mathbf{T}_c-\mathbf{T}_{\overline{c}}) (\mathbf{T}_c-\mathbf{T}_{\overline{c}}) | ^1S_0^{[c]}, \mathrm{Born} \rangle.
\label{eq:OpRenTransformed1P1Two}
\end{eqnarray}

Finally, we derive the formula for the running of the LDME
$\langle {\cal O}^{H}[^3S_1^{[8]}]\rangle^{(\mu_\Lambda)}$ with $\mu_\Lambda$.
Differentiating Eq.~(\ref{eq:NLOLDME2}) with respect to $\mu_\Lambda$, we obtain
a renormalization group equation, with the solution
\begin{eqnarray}
 \langle {\cal O}^{H}[^3S_1^{[8]}]\rangle^{(\mu_\Lambda)} &=& \langle {\cal O}^{H}[^3S_1^{[8]}]\rangle^{(\mu_{\Lambda,0})} + \frac{16}{3m_Q^2} \left(\frac{1}{\beta_0}\ln \frac{\alpha_s(\mu_{\Lambda,0})}{\alpha_s(\mu_\Lambda)}+\frac{\beta_1}{4\pi\beta_0^2}(\alpha_s(\mu_\Lambda)-\alpha_s(\mu_{\Lambda,0}))\right) \nonumber
 \\
 &&\times \sum_J
\left[ \frac{C_F}{2C_A} \langle {\cal O}^{H}[^3P_J^{[1]}]\rangle
 + \frac{C_A^2-4}{4C_A} \langle {\cal O}^{H}[^3P_J^{[8]}]\rangle \right],
\end{eqnarray}
through NLO in $\alpha_s$. Here, $\beta_0=\frac{11}{3}C_A-\frac{2}{3}n_f$ and $\beta_1=\frac{34}{3}C_A^2-2C_Fn_f - \frac{10}{3}C_An_f$. The evolution equations for $\langle {\cal O}^{H}[^3S_1^{[1]}]\rangle^{(\mu_\Lambda)}$, $\langle {\cal O}^{H}[^1S_0^{[8]}]\rangle^{(\mu_\Lambda)}$, and $\langle {\cal O}^{H}[^1S_0^{[1]}]\rangle^{(\mu_\Lambda)}$ may be obtained similarly and read
\begin{eqnarray}
\langle {\cal O}^{H}[^3S_1^{[1]}]\rangle^{(\mu_\Lambda)} &=& \langle {\cal O}^{H}[^3S_1^{[1]}]\rangle^{(\mu_{\Lambda,0})} + \frac{16}{3m_Q^2} \left(\frac{1}{\beta_0}\ln \frac{\alpha_s(\mu_{\Lambda,0})}{\alpha_s(\mu_\Lambda)}+\frac{\beta_1}{4\pi\beta_0^2}(\alpha_s(\mu_\Lambda)-\alpha_s(\mu_{\Lambda,0}))\right) \nonumber
 \\
 &&\times \sum_J \langle {\cal O}^{H}[^3P_J^{[8]}]\rangle,
\\
\langle {\cal O}^{H}[^1S_0^{[8]}]\rangle^{(\mu_\Lambda)} &=& \langle {\cal O}^{H}[^1S_0^{[8]}]\rangle^{(\mu_{\Lambda,0})} + \frac{16}{3m_Q^2} \left(\frac{1}{\beta_0}\ln \frac{\alpha_s(\mu_{\Lambda,0})}{\alpha_s(\mu_\Lambda)}+\frac{\beta_1}{4\pi\beta_0^2}(\alpha_s(\mu_\Lambda)-\alpha_s(\mu_{\Lambda,0}))\right) \nonumber
 \\
 &&\times 
\left[ \frac{C_F}{2C_A} \langle {\cal O}^{H}[^1P_1^{[1]}]\rangle
 + \frac{C_A^2-4}{4C_A} \langle {\cal O}^{H}[^1P_1^{[8]}]\rangle \right],
\\
\langle {\cal O}^{H}[^1S_0^{[1]}]\rangle^{(\mu_\Lambda)} &=& \langle {\cal O}^{H}[^1S_0^{[1]}]\rangle^{(\mu_{\Lambda,0})} + \frac{16}{3m_Q^2} \left(\frac{1}{\beta_0}\ln \frac{\alpha_s(\mu_{\Lambda,0})}{\alpha_s(\mu_\Lambda)}+\frac{\beta_1}{4\pi\beta_0^2}(\alpha_s(\mu_\Lambda)-\alpha_s(\mu_{\Lambda,0}))\right) \nonumber
 \\
 &&\times \langle {\cal O}^{H}[^1P_1^{[8]}]\rangle.
\end{eqnarray}

\section{Dipole subtraction for quarkonium production\label{sec:Subtraction}}

\subsection{General setup}

In a preliminary version, not yet taking into account kinematic cuts, we write the partonic NLO corrections as
\begin{eqnarray}
 \int d\hat{\sigma} &=& \int d\mathrm{PS}_3 \frac{d\hat{\sigma}_\mathrm{real}}{d\mathrm{PS}_3} + \int d\mathrm{PS}_2 \frac{d\hat{\sigma}_\mathrm{virtual}+d\hat{\sigma}_\mathrm{MFC}+d\hat{\sigma}_\mathrm{op.\,ren.}}{d\mathrm{PS}_2} \nonumber \\
 &=& \int d\mathrm{PS}_3 \left( \frac{d\hat{\sigma}_\mathrm{real}}{d\mathrm{PS}_3} - \frac{d\hat{\sigma}_\mathrm{subtr}}{d\mathrm{PS}_3} \right) \nonumber \\
 &&+ \int d\mathrm{PS}_2 \left( \frac{d\hat{\sigma}_\mathrm{virtual}+d\hat{\sigma}_\mathrm{MFC}+d\hat{\sigma}_\mathrm{op.\,ren.}}{d\mathrm{PS}_2} +  [dx] \int d\mathrm{PS}_\mathrm{dipole} \frac{d\hat{\sigma}_\mathrm{subtr}}{d\mathrm{PS}_3} \right). \label{eq:dipolesubgeneral}
\end{eqnarray}
Here, $d\mathrm{PS}_2$ is the two-particle phase space element, and $d\mathrm{PS}_3$ is the three-particle phase space element, which factorizes in some way, as either $d\mathrm{PS}_3=d\mathrm{PS}_2 d\mathrm{PS}_\mathrm{dipole}$ or $d\mathrm{PS}_3=d\mathrm{PS}_2 dx\, d\mathrm{PS}_\mathrm{dipole}$, where $d\mathrm{PS}_\mathrm{dipole}$ are
certain {\em dipole phase space elements} and $dx$ matches its counterpart within $d\hat\sigma_\mathrm{MFC}$ as defined in Eq.~(\ref{eq:MFC}). The {\em subtraction terms} $d\hat\sigma_\mathrm{subtr}$ are defined in terms of some kinematic variables in the parameterization of $d\mathrm{PS}_\mathrm{dipole}$ and certain $2\to2$ kinematics momenta $\{\tilde{p}_i\}$ appearing in $d\mathrm{PS}_2$, which are in turn in some way {\em mapped} onto the $2\to3$ kinematics momenta $\{p_i\}$. The idea is that $d\hat\sigma_\mathrm{subtr}$ matches $d\hat\sigma_\mathrm{real}$ in all singular limits. Therefore, the first bracket on the right-hand side of Eq.~(\ref{eq:dipolesubgeneral}) is free of divergences and can be integrated numerically in four dimensions. On the other hand, $d\hat\sigma_\mathrm{subtr}$ is simple enough that it can be analytically integrated in $D$ dimensions over $d\mathrm{PS}_\mathrm{dipole}$. The IR poles of $d\sigma_\mathrm{dipole}$ then become explicit as $\epsilon^{-1}$ and $\epsilon^{-2}$ poles and cancel the singularities of $d\sigma_\mathrm{virtual}+d\hat{\sigma}_\mathrm{MFC}+d\hat{\sigma}_\mathrm{op.\,ren.}$, so that the second bracket on the right-hand side of Eq.~(\ref{eq:dipolesubgeneral}) is also finite and can be integrated numerically over $d\mathrm{PS}_2$ or $d\mathrm{PS}_2dx$ in four dimensions, too. So the task is to construct appropriate expressions for $d\sigma_\mathrm{subtr}$ and $d\mathrm{PS}_\mathrm{dipole}$ with the corresponding momentum mappings.

\subsection{Subtraction term}

\begin{table}
\begin{center}
\begin{tabular}{l|cc|l|l}
& $p_i$ & $p_k$ & Definition and Integration & Applied Mapping
\\
\hline
 $V^{\mathrm{ini,}S_1}_{ij,k}$ &$p_1$ or $p_2$ & $p_0$ & PW, section 6.1 & MapPW6($p_i, p_j$)\\
 $V^{\mathrm{ini,}S_1}_{ij,k}$ &$p_1$ or $p_2$ & $p_1$ or $p_2$ & CS, section 5.6 ($n=p_3+p_4$)& MapCS($p_i$) \\
 $V^{\mathrm{ini,}S_1}_{ij,k}$ &$p_1$ or $p_2$ & $p_3$ or $p_4$ & CS, section 5.3 & MapCS($p_i$) \\
 $V^{\mathrm{fin,}S_1}_{ij,k}$ &$p_0$ & $p_1$ or $p_2$ & PW, section 6.2 & MapPW6($p_k, p_j$) \\
 $V^{\mathrm{fin,}S_1}_{ij,k}$&$p_0$ & $p_3$ or $p_4$ & PW, section 5.2 & MapPW5.2($p_j$) \\
 $V^{\mathrm{fin,}S_1}_{ij,k}$&$p_3$ or $p_4$ & $p_0$ & PW, section 5.1 & MapPW5.1($p_i$)\\
 $V^{\mathrm{fin,}S_1}_{ij,k}$&$p_3$ or $p_4$ & $p_1$ or $p_2$ & CS, section 5.2 & MapCS($p_k$) \\
 $V_{S_2,ij}$ & $p_1$ or $p_2$ && Here, Eq.~(\ref{eq:S2Vterm}) and section \ref{sec:S2dipolesini} & MapPW6($p_i,p_j$) \\
 $V_{S_2,ij}$ & $p_3$ or $p_4$ && Here, Eq.~(\ref{eq:S2Vterm}) and section \ref{sec:S2dipolesfin} & MapPW5.2($p_j$) \\
 $V_{S_3,j}$ && & Here, Eq.~(\ref{eq:S3Vterm}) and section \ref{sec:S3dipoles} & MapPW5.2($p_j$)
\end{tabular}
\end{center}
\caption{\label{tab:DipoleOverview}List of occurring $V$ terms with given momentum assignments; of where their definitions and analytic expressions upon integration over the dipole phase spaces may be found in the Catani-Seymour (CS) \cite{Catani:1996vz} and Phaf-Weinzierl (PW) \cite{Phaf:2001gc} papers and here; and of momentum mappings, according to the naming scheme of section~\ref{sec:MomentumMapping}, to be applied to the numerical integrations of the respective dipole terms over $d\mathrm{PS}_3$.}
\end{table}

From Eqs.~(\ref{eq:soft1singular}), (\ref{eq:IniColl}), and (\ref{eq:FinColl}), we observe that the sum of all softly and collinearly divergent terms can be brought into a form that can be approximated in all singular limits by the subtraction term
\begin{eqnarray}
 \frac{d\hat{\sigma}_\mathrm{subtr}(a+b\to Q\overline{Q}[n]+X)}{d\mathrm{PS}_3} &=& \frac{1}{N_\mathrm{col}(n) N_\mathrm{pol}(n)}\,\frac{1}{2(p_1+p_2)^2} \nonumber \\
 &&\times \frac{F_\mathrm{sym}(X)}{n_\mathrm{col}(a) n_\mathrm{pol}(a) n_\mathrm{col}(b) n_\mathrm{pol}(b)} \, \| | abn, \mathrm{subtr}\rangle \|^2,
 \label{eq:dipolesmoredetailedOne}
\end{eqnarray}
with
\begin{eqnarray}
 \| | abn, \mathrm{subtr} \rangle \|^2 &=& \sum_{j=3}^4 \sum_{i=1}^2 \sum_{\substack{k=0\\k\neq i,j}}^4
 \frac{n_\mathrm{col}(i)}{n_\mathrm{col}((ij))}
  \frac{-1}{2p_i\cdot p_j} \frac{1}{x} \langle n, \mathrm{Born} | V^{\mathrm{ini,}S_1}_{ij,k}\frac{\mathbf{T}_{(ij)} \mathbf{T}_k}{\mathbf{T}_{(ij)}^2} | n, \mathrm{Born} \rangle 
\nonumber \\
&& + \sum_{j=3}^4 \sum_{\substack{i=0 \\ i\neq 1,2,j}}^3 \sum_{\substack{k=0\\k\neq i,j}}^4 \frac{-1}{2p_i\cdot p_j} \langle n, \mathrm{Born} | V^{\mathrm{fin,}S_1}_{ij,k}\frac{\mathbf{T}_{(ij)} \mathbf{T}_k}{\mathbf{T}_{(ij)}^2} | n, \mathrm{Born} \rangle
\begin{cases}
\frac{1}{x} & \mathrm{if}\,k=1,2 \\ 1 & \mathrm{if}\,k\neq1,2
\end{cases}
\nonumber \\
&&+ \sum_{j=3}^4 \sum_{\substack{i=1 \\ i\neq j}}^4  V_{S_2,ij}^{\beta} \epsilon_\beta(m_l) \langle n,\mathrm{Born} | \mathbf{T}_{(ij)} (\mathbf{T}_c-\mathbf{T}_{\overline{c}}) | m(n),\mathrm{Born}\rangle
\nonumber \\
&&+ \sum_{j=3}^4 V_{S_3,j}^{\alpha\beta} \epsilon^\ast_\alpha(m_l)\epsilon_\beta(m_l) \langle m(n),\mathrm{Born} |  (\mathbf{T}_c-\mathbf{T}_{\overline{c}}) (\mathbf{T}_c-\mathbf{T}_{\overline{c}}) | m(n),\mathrm{Born}\rangle.\qquad
\label{eq:dipolesmoredetailed}
\end{eqnarray}
In the respective limits, the initial-state collinear singularities are reproduced by the first line, the final-state collinear singularities by the second line, the $S_1$ soft divergences by the corresponding soft limits of the first and second lines together, and the $S_2$ and $S_3$ soft divergences in the case of $P$-wave states by the last two lines. In the regions away from the soft and collinear limits, there are no additional singularities. We call each of the terms in the sums a {\em dipole}. In the Born amplitudes, particles $i$ and $j$ are replaced by one particle $(ij)$, which is a gluon, a light quark or the $Q\overline{Q}[n]$ state depending on the collinear or soft limits to be approximated. Where there is no divergent collinear or soft limit to be approximated, the contribution is just zero. We note that, in the soft limits, particles $i$ and $(ij)$ are the same and that, in the soft and final-state collinear limits, $x=1$. We further define $m({^3P}_J^{[1/8]})={^3S}_1^{[1/8]}$ and $m({^1P}_1^{[1/8]})={^1S}_0^{[1/8]}$. Table~\ref{tab:DipoleOverview} lists where to find the explicit expressions for $V^{\mathrm{ini,}S_1}_{ij,k}$ and $V^{\mathrm{fin,}S_1}_{ij,k}$ in the Catani-Seymour \cite{Catani:1996vz} and Phaf-Weinzierl \cite{Phaf:2001gc} papers. The factors in Eq.~(\ref{eq:dipolesmoredetailed}) are adjusted so that $V^{\mathrm{ini,}S_1}_{ij,k}$ equals $V^{ij}_k$ or $V^{ij,k}$ and $V^{\mathrm{fin,}S_1}_{ij,k}$ equals $V_{ij,k}$ or $V_{ij}^k$ in their notations. The particle $(ij)$ is called an {\em emitter}, the particle $k$ a {\em spectator}, and the indices $s$ and $s'$ or $\mu$ and $\nu$ within $V_{ij,k}$ are the spin or polarization indices of particle $(ij)$ in the Born amplitude. The $V_{S_2,ij}^{\beta}$ and $V_{S_3,j}^{\alpha\beta}$ terms, which are new, are given by
\begin{eqnarray}
 V_{S_2,ij}^\beta &=& 4 g_s^2 \left( -\frac{p_i^\beta}{p_i\cdot p_j\; p_0\cdot p_j} + \frac{p_0\cdot p_i p_j^\beta}{p_i\cdot p_j (p_0\cdot p_j)^2} -\frac{p_0^2 p_j^\beta}{(p_0\cdot p_j)^3} \right), \label{eq:S2Vterm}\\
 V_{S_3,j}^{\alpha\beta} &=& 4 g_s^2 \left( - \frac{g^{\alpha\beta}}{(p_0\cdot p_j)^2} - \frac{p_0^2 p_j^\alpha p_j^\beta}{(p_0\cdot p_j)^4} \right), \label{eq:S3Vterm}
\end{eqnarray}
so as to approximate Eqs.~(\ref{eq:s2term}) and (\ref{eq:s3term}). A pictorial summary of all dipole terms appearing in our study is given in Fig.~\ref{fig:listofdipoles}.

\newcommand{\TODcolwidth}{119pt}
\newcommand{\TODscalefactor}{0.6}
\newcommand{\TODhspace}{0pt}
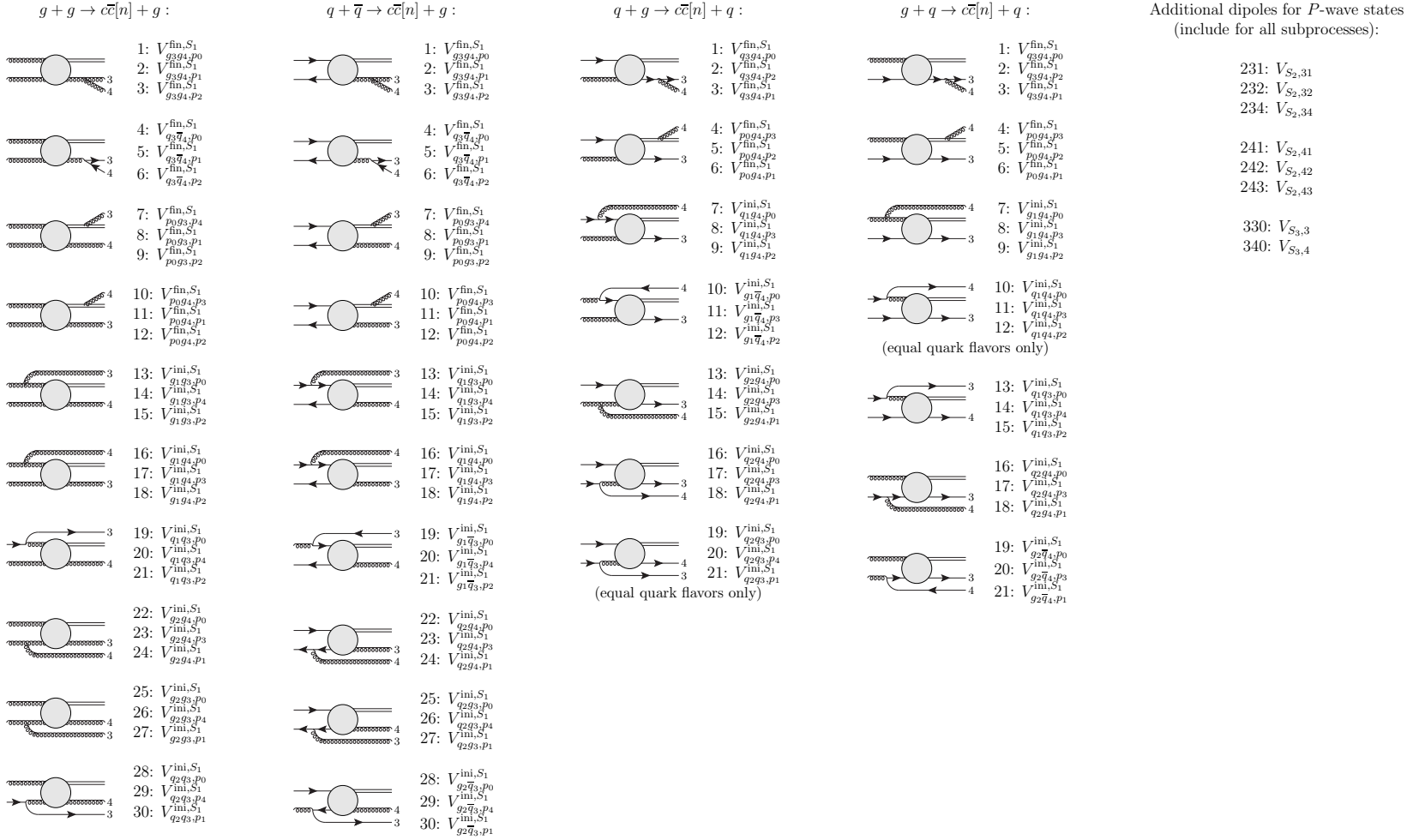
\begin{sidewaysfigure}
\begin{center}
\begin{minipage}[t]{\TODcolwidth}
\vspace{0pt}
\scalebox{\TODscalefactor}{
\begin{tabular}{cc}
\multicolumn{2}{c}{\bf $g+g\to c\overline{c}[n] + g:$} \\
\\
\multirow{2}{*}{
\begin{picture}(77,44)(0,0)
\Line(36,30)(72,30)
\Line(36,28)(72,28)
\Gluon(0,29)(36,29){1.5}{12}
\Gluon(0,15)(36,15){1.5}{12}
\Gluon(36,15)(72,15){1.5}{12}
\Gluon(56,15)(72,6){1.5}{6}
\GOval(36,22)(11,11)(0){0.9}
\Text(74,15)[l]{\scriptsize 3}
\Text(74,6)[l]{\scriptsize 4}
\end{picture}
}
& 1: $V_{g_3g_4,p_0}^{\mathrm{fin,}S_1}$ \\
& 2: $V_{g_3g_4,p_1}^{\mathrm{fin,}S_1}$ \\
& 3: $V_{g_3g_4,p_2}^{\mathrm{fin,}S_1}$ \\
\\
\multirow{2}{*}{
\begin{picture}(77,44)(0,0)
\Line(36,30)(72,30)
\Line(36,28)(72,28)
\Gluon(0,29)(36,29){1.5}{12}
\Gluon(0,15)(36,15){1.5}{12}
\Gluon(36,15)(57,15){1.5}{6}
\ArrowLine(57,15)(72,15)
\ArrowLine(72,6)(57,15)
\GOval(36,22)(11,11)(0){0.9}
\Text(74,15)[l]{\scriptsize 3}
\Text(74,6)[l]{\scriptsize 4}
\end{picture}
}
& 4: $V_{q_3\overline{q}_4,p_0}^{\mathrm{fin,}S_1}$ \\
& 5: $V_{q_3\overline{q}_4,p_1}^{\mathrm{fin,}S_1}$ \\
& 6: $V_{q_3\overline{q}_4,p_2}^{\mathrm{fin,}S_1}$ \\
\\
\multirow{2}{*}{
\begin{picture}(77,44)(0,0)
\Line(36,30)(72,30)
\Line(36,28)(72,28)
\Gluon(0,29)(36,29){1.5}{12}
\Gluon(0,15)(36,15){1.5}{12}
\Gluon(36,15)(72,15){1.5}{12}
\Gluon(57,29)(72,38){1.5}{6}
\GOval(36,22)(11,11)(0){0.9}
\Text(74,38)[l]{\scriptsize 3}
\Text(74,15)[l]{\scriptsize 4}
\end{picture}
}
& 7: $V_{p_0g_3,p_4}^{\mathrm{fin,}S_1}$ \\
& 8: $V_{p_0g_3,p_1}^{\mathrm{fin,}S_1}$ \\
& 9: $V_{p_0g_3,p_2}^{\mathrm{fin,}S_1}$ \\
\\
\multirow{2}{*}{
\begin{picture}(77,44)(0,0)
\Line(36,30)(72,30)
\Line(36,28)(72,28)
\Gluon(0,29)(36,29){1.5}{12}
\Gluon(0,15)(36,15){1.5}{12}
\Gluon(36,15)(72,15){1.5}{12}
\Gluon(57,29)(72,38){1.5}{6}
\GOval(36,22)(11,11)(0){0.9}
\Text(74,38)[l]{\scriptsize 4}
\Text(74,15)[l]{\scriptsize 3}
\end{picture}
}
& 10: $V_{p_0g_4,p_3}^{\mathrm{fin,}S_1}$ \\
& 11: $V_{p_0g_4,p_1}^{\mathrm{fin,}S_1}$ \\
& 12: $V_{p_0g_4,p_2}^{\mathrm{fin,}S_1}$ \\
\\
\multirow{2}{*}{
\begin{picture}(77,44)(0,0)
\Line(36,30)(72,30)
\Line(36,28)(72,28)
\Gluon(0,29)(36,29){1.5}{12}
\Gluon(0,15)(36,15){1.5}{12}
\Gluon(36,15)(72,15){1.5}{12}
\Gluon(20,38)(72,38){1.5}{18}
\GlueArc(23,29)(9,90,180){1.5}{4}
\GOval(36,22)(11,11)(0){0.9}
\Text(74,38)[l]{\scriptsize 3}
\Text(74,15)[l]{\scriptsize 4}
\end{picture}
}
& 13: $V_{g_1g_3,p_0}^{\mathrm{ini,}S_1}$ \\
& 14: $V_{g_1g_3,p_4}^{\mathrm{ini,}S_1}$ \\
& 15: $V_{g_1g_3,p_2}^{\mathrm{ini,}S_1}$ \\
\\
\multirow{2}{*}{
\begin{picture}(77,44)(0,0)
\Line(36,30)(72,30)
\Line(36,28)(72,28)
\Gluon(0,29)(36,29){1.5}{12}
\Gluon(0,15)(36,15){1.5}{12}
\Gluon(36,15)(72,15){1.5}{12}
\Gluon(20,38)(72,38){1.5}{18}
\GlueArc(23,29)(9,90,180){1.5}{4}
\GOval(36,22)(11,11)(0){0.9}
\Text(74,38)[l]{\scriptsize 4}
\Text(74,15)[l]{\scriptsize 3}
\end{picture}
}
& 16: $V_{g_1g_4,p_0}^{\mathrm{ini,}S_1}$ \\
& 17: $V_{g_1g_4,p_3}^{\mathrm{ini,}S_1}$ \\
& 18: $V_{g_1g_4,p_2}^{\mathrm{ini,}S_1}$ \\
\\
\multirow{2}{*}{
\begin{picture}(77,44)(0,0)
\Line(36,30)(72,30)
\Line(36,28)(72,28)
\Gluon(14,29)(36,29){1.5}{7}
\Gluon(0,15)(36,15){1.5}{12}
\Gluon(36,15)(72,15){1.5}{12}
\ArrowLine(23,38)(72,38)
\CArc(23,29)(9,90,180)
\ArrowLine(0,29)(14,29)
\GOval(36,22)(11,11)(0){0.9}
\Text(74,38)[l]{\scriptsize 3}
\Text(74,15)[l]{\scriptsize 4}
\end{picture}
}
& 19: $V_{q_1q_3,p_0}^{\mathrm{ini,}S_1}$ \\
& 20: $V_{q_1q_3,p_4}^{\mathrm{ini,}S_1}$ \\
& 21: $V_{q_1q_3,p_2}^{\mathrm{ini,}S_1}$ \\
\\
\multirow{2}{*}{
\begin{picture}(77,44)(0,0)
\Line(36,30)(72,30)
\Line(36,28)(72,28)
\Gluon(0,29)(36,29){1.5}{12}
\Gluon(0,15)(36,15){1.5}{12}
\Gluon(36,15)(72,15){1.5}{12}
\Gluon(20,6)(72,6){1.5}{18}
\GlueArc(23,15)(9,180,270){-1.5}{4}
\GOval(36,22)(11,11)(0){0.9}
\Text(74,15)[l]{\scriptsize 3}
\Text(74,6)[l]{\scriptsize 4}
\end{picture}
}
& 22: $V_{g_2g_4,p_0}^{\mathrm{ini,}S_1}$ \\
& 23: $V_{g_2g_4,p_3}^{\mathrm{ini,}S_1}$ \\
& 24: $V_{g_2g_4,p_1}^{\mathrm{ini,}S_1}$ \\
\\
\multirow{2}{*}{
\begin{picture}(77,44)(0,0)
\Line(36,30)(72,30)
\Line(36,28)(72,28)
\Gluon(0,29)(36,29){1.5}{12}
\Gluon(0,15)(36,15){1.5}{12}
\Gluon(36,15)(72,15){1.5}{12}
\Gluon(20,6)(72,6){1.5}{18}
\GlueArc(23,15)(9,180,270){-1.5}{4}
\GOval(36,22)(11,11)(0){0.9}
\Text(74,15)[l]{\scriptsize 4}
\Text(74,6)[l]{\scriptsize 3}
\end{picture}
}
& 25: $V_{g_2g_3,p_0}^{\mathrm{ini,}S_1}$ \\
& 26: $V_{g_2g_3,p_4}^{\mathrm{ini,}S_1}$ \\
& 27: $V_{g_2g_3,p_1}^{\mathrm{ini,}S_1}$ \\
\\
\multirow{2}{*}{
\begin{picture}(77,44)(0,0)
\Line(36,30)(72,30)
\Line(36,28)(72,28)
\Gluon(0,29)(36,29){1.5}{12}
\Gluon(14,15)(36,15){1.5}{7}
\Gluon(36,15)(72,15){1.5}{12}
\ArrowLine(0,15)(14,15)
\CArc(23,15)(9,180,270)
\ArrowLine(23,6)(72,6)
\GOval(36,22)(11,11)(0){0.9}
\Text(74,15)[l]{\scriptsize 4}
\Text(74,6)[l]{\scriptsize 3}
\end{picture}
}
& 28: $V_{q_2q_3,p_0}^{\mathrm{ini,}S_1}$ \\
& 29: $V_{q_2q_3,p_4}^{\mathrm{ini,}S_1}$ \\
& 30: $V_{q_2q_3,p_1}^{\mathrm{ini,}S_1}$
\end{tabular}
}
\end{minipage}
\hspace{\TODhspace}
\begin{minipage}[t]{\TODcolwidth}
\vspace{0pt}
\scalebox{\TODscalefactor}{
\begin{tabular}{cc}
\multicolumn{2}{c}{\bf $q+\overline{q}\to c\overline{c}[n] + g:$} \\
\\
\multirow{2}{*}{
\begin{picture}(77,44)(0,0)
\Line(36,30)(72,30)
\Line(36,28)(72,28)
\ArrowLine(0,29)(28,29)
\ArrowLine(28,15)(0,15)
\Gluon(36,15)(72,15){1.5}{12}
\Gluon(56,15)(72,6){1.5}{6}
\GOval(36,22)(11,11)(0){0.9}
\Text(74,15)[l]{\scriptsize 3}
\Text(74,6)[l]{\scriptsize 4}
\end{picture}
}
& 1: $V_{g_3g_4,p_0}^{\mathrm{fin,}S_1}$ \\
& 2: $V_{g_3g_4,p_1}^{\mathrm{fin,}S_1}$ \\
& 3: $V_{g_3g_4,p_2}^{\mathrm{fin,}S_1}$ \\
\\
\multirow{2}{*}{
\begin{picture}(77,44)(0,0)
\Line(36,30)(72,30)
\Line(36,28)(72,28)
\ArrowLine(0,29)(28,29)
\ArrowLine(28,15)(0,15)
\Gluon(36,15)(57,15){1.5}{6}
\ArrowLine(57,15)(72,15)
\ArrowLine(72,6)(57,15)
\GOval(36,22)(11,11)(0){0.9}
\Text(74,15)[l]{\scriptsize 3}
\Text(74,6)[l]{\scriptsize 4}
\end{picture}
}
& 4: $V_{q_3\overline{q}_4,p_0}^{\mathrm{fin,}S_1}$ \\
& 5: $V_{q_3\overline{q}_4,p_1}^{\mathrm{fin,}S_1}$ \\
& 6: $V_{q_3\overline{q}_4,p_2}^{\mathrm{fin,}S_1}$ \\
\\
\multirow{2}{*}{
\begin{picture}(77,44)(0,0)
\Line(36,30)(72,30)
\Line(36,28)(72,28)
\ArrowLine(0,29)(28,29)
\ArrowLine(28,15)(0,15)
\Gluon(36,15)(72,15){1.5}{12}
\Gluon(57,29)(72,38){1.5}{6}
\GOval(36,22)(11,11)(0){0.9}
\Text(74,38)[l]{\scriptsize 3}
\Text(74,15)[l]{\scriptsize 4}
\end{picture}
}
& 7: $V_{p_0g_3,p_4}^{\mathrm{fin,}S_1}$ \\
& 8: $V_{p_0g_3,p_1}^{\mathrm{fin,}S_1}$ \\
& 9: $V_{p_0g_3,p_2}^{\mathrm{fin,}S_1}$ \\
\\
\multirow{2}{*}{
\begin{picture}(77,44)(0,0)
\Line(36,30)(72,30)
\Line(36,28)(72,28)
\ArrowLine(0,29)(28,29)
\ArrowLine(28,15)(0,15)
\Gluon(36,15)(72,15){1.5}{12}
\Gluon(57,29)(72,38){1.5}{6}
\GOval(36,22)(11,11)(0){0.9}
\Text(74,38)[l]{\scriptsize 4}
\Text(74,15)[l]{\scriptsize 3}
\end{picture}
}
& 10: $V_{p_0g_4,p_3}^{\mathrm{fin,}S_1}$ \\
& 11: $V_{p_0g_4,p_1}^{\mathrm{fin,}S_1}$ \\
& 12: $V_{p_0g_4,p_2}^{\mathrm{fin,}S_1}$ \\
\\
\multirow{2}{*}{
\begin{picture}(77,44)(0,0)
\Line(36,30)(72,30)
\Line(36,28)(72,28)
\ArrowLine(0,29)(14,29)
\ArrowLine(14,29)(28,29)
\ArrowLine(28,15)(0,15)
\Gluon(36,15)(72,15){1.5}{12}
\Gluon(20,38)(72,38){1.5}{18}
\GlueArc(23,29)(9,90,180){1.5}{4}
\GOval(36,22)(11,11)(0){0.9}
\Text(74,38)[l]{\scriptsize 3}
\Text(74,15)[l]{\scriptsize 4}
\end{picture}
}
& 13: $V_{q_1g_3,p_0}^{\mathrm{ini,}S_1}$ \\
& 14: $V_{q_1g_3,p_4}^{\mathrm{ini,}S_1}$ \\
& 15: $V_{q_1g_3,p_2}^{\mathrm{ini,}S_1}$ \\
\\
\multirow{2}{*}{
\begin{picture}(77,44)(0,0)
\Line(36,30)(72,30)
\Line(36,28)(72,28)
\ArrowLine(0,29)(14,29)
\ArrowLine(14,29)(28,29)
\ArrowLine(28,15)(0,15)
\Gluon(36,15)(72,15){1.5}{12}
\Gluon(20,38)(72,38){1.5}{18}
\GlueArc(23,29)(9,90,180){1.5}{4}
\GOval(36,22)(11,11)(0){0.9}
\Text(74,38)[l]{\scriptsize 4}
\Text(74,15)[l]{\scriptsize 3}
\end{picture}
}
& 16: $V_{q_1g_4,p_0}^{\mathrm{ini,}S_1}$ \\
& 17: $V_{q_1g_4,p_3}^{\mathrm{ini,}S_1}$ \\
& 18: $V_{q_1g_4,p_2}^{\mathrm{ini,}S_1}$ \\
\\
\multirow{2}{*}{
\begin{picture}(77,44)(0,0)
\Line(36,30)(72,30)
\Line(36,28)(72,28)
\ArrowLine(14,29)(28,29)
\Gluon(0,29)(14,29){1.5}{4}
\ArrowLine(28,15)(0,15)
\Gluon(36,15)(72,15){1.5}{12}
\ArrowLine(72,38)(23,38)
\CArc(23,29)(9,90,180)
\GOval(36,22)(11,11)(0){0.9}
\Text(74,38)[l]{\scriptsize 3}
\Text(74,15)[l]{\scriptsize 4}
\end{picture}
}
& 19: $V_{g_1\overline{q}_3,p_0}^{\mathrm{ini,}S_1}$ \\
& 20: $V_{g_1\overline{q}_3,p_4}^{\mathrm{ini,}S_1}$ \\
& 21: $V_{g_1\overline{q}_3,p_2}^{\mathrm{ini,}S_1}$ \\
\\
\multirow{2}{*}{
\begin{picture}(77,44)(0,0)
\Line(36,30)(72,30)
\Line(36,28)(72,28)
\ArrowLine(0,29)(28,29)
\ArrowLine(28,15)(14,15)
\ArrowLine(14,15)(0,15)
\Gluon(36,15)(72,15){1.5}{12}
\Gluon(20,6)(72,6){1.5}{18}
\GlueArc(23,15)(9,180,270){-1.5}{4}
\GOval(36,22)(11,11)(0){0.9}
\Text(74,15)[l]{\scriptsize 3}
\Text(74,6)[l]{\scriptsize 4}
\end{picture}
}
& 22: $V_{q_2g_4,p_0}^{\mathrm{ini,}S_1}$ \\
& 23: $V_{q_2g_4,p_3}^{\mathrm{ini,}S_1}$ \\
& 24: $V_{q_2g_4,p_1}^{\mathrm{ini,}S_1}$ \\
\\
\multirow{2}{*}{
\begin{picture}(77,44)(0,0)
\Line(36,30)(72,30)
\Line(36,28)(72,28)
\ArrowLine(0,29)(28,29)
\ArrowLine(28,15)(14,15)
\ArrowLine(14,15)(0,15)
\Gluon(36,15)(72,15){1.5}{12}
\Gluon(20,6)(72,6){1.5}{18}
\GlueArc(23,15)(9,180,270){-1.5}{4}
\GOval(36,22)(11,11)(0){0.9}
\Text(74,15)[l]{\scriptsize 4}
\Text(74,6)[l]{\scriptsize 3}
\end{picture}
}
& 25: $V_{q_2g_3,p_0}^{\mathrm{ini,}S_1}$ \\
& 26: $V_{q_2g_3,p_4}^{\mathrm{ini,}S_1}$ \\
& 27: $V_{q_2g_3,p_1}^{\mathrm{ini,}S_1}$ \\
\\
\multirow{2}{*}{
\begin{picture}(77,44)(0,0)
\Line(36,30)(72,30)
\Line(36,28)(72,28)
\ArrowLine(0,29)(28,29)
\ArrowLine(28,15)(14,15)
\Gluon(36,15)(72,15){1.5}{12}
\Gluon(0,15)(14,15){1.5}{4}
\CArc(23,15)(9,180,270)
\ArrowLine(23,6)(72,6)
\GOval(36,22)(11,11)(0){0.9}
\Text(74,15)[l]{\scriptsize 4}
\Text(74,6)[l]{\scriptsize 3}
\end{picture}
}
& 28: $V_{g_2\overline{q}_3,p_0}^{\mathrm{ini,}S_1}$ \\
& 29: $V_{g_2\overline{q}_3,p_4}^{\mathrm{ini,}S_1}$ \\
& 30: $V_{g_2\overline{q}_3,p_1}^{\mathrm{ini,}S_1}$
\end{tabular}
}
\end{minipage}
\hspace{\TODhspace}
\begin{minipage}[t]{\TODcolwidth}
\vspace{0pt}
\scalebox{\TODscalefactor}{
\begin{tabular}{cc}
\multicolumn{2}{c}{\bf $q+g\to c\overline{c}[n] + q:$} \\
\\
\multirow{2}{*}{
\begin{picture}(77,44)(0,0)
\Line(36,30)(72,30)
\Line(36,28)(72,28)
\ArrowLine(0,29)(28,29)
\Gluon(0,15)(36,15){1.5}{12}
\ArrowLine(42,15)(60,15)
\ArrowLine(60,15)(72,15)
\Gluon(56,15)(72,6){1.5}{6}
\GOval(36,22)(11,11)(0){0.9}
\Text(74,15)[l]{\scriptsize 3}
\Text(74,6)[l]{\scriptsize 4}
\end{picture}
}
& 1: $V_{q_3g_4,p_0}^{\mathrm{fin,}S_1}$ \\
& 2: $V_{q_3g_4,p_2}^{\mathrm{fin,}S_1}$ \\
& 3: $V_{q_3g_4,p_1}^{\mathrm{fin,}S_1}$ \\
\\
\multirow{2}{*}{
\begin{picture}(77,44)(0,0)
\Line(36,30)(72,30)
\Line(36,28)(72,28)
\ArrowLine(0,29)(28,29)
\Gluon(0,15)(36,15){1.5}{12}
\ArrowLine(42,15)(72,15)
\Gluon(57,29)(72,38){1.5}{6}
\GOval(36,22)(11,11)(0){0.9}
\Text(74,38)[l]{\scriptsize 4}
\Text(74,15)[l]{\scriptsize 3}
\end{picture}
}
& 4: $V_{p_0g_4,p_3}^{\mathrm{fin,}S_1}$ \\
& 5: $V_{p_0g_4,p_2}^{\mathrm{fin,}S_1}$ \\
& 6: $V_{p_0g_4,p_1}^{\mathrm{fin,}S_1}$ \\
\\
\multirow{2}{*}{
\begin{picture}(77,44)(0,0)
\Line(36,30)(72,30)
\Line(36,28)(72,28)
\ArrowLine(0,29)(14,29)
\ArrowLine(14,29)(28,29)
\Gluon(0,15)(36,15){1.5}{12}
\ArrowLine(42,15)(72,15)
\Gluon(20,38)(72,38){1.5}{18}
\GlueArc(23,29)(9,90,180){1.5}{4}
\GOval(36,22)(11,11)(0){0.9}
\Text(74,38)[l]{\scriptsize 4}
\Text(74,15)[l]{\scriptsize 3}
\end{picture}
}
& 7: $V_{q_1g_4,p_0}^{\mathrm{ini,}S_1}$ \\
& 8: $V_{q_1g_4,p_3}^{\mathrm{ini,}S_1}$ \\
& 9: $V_{q_1g_4,p_2}^{\mathrm{ini,}S_1}$ \\
\\
\multirow{2}{*}{
\begin{picture}(77,44)(0,0)
\Line(36,30)(72,30)
\Line(36,28)(72,28)
\Gluon(0,15)(36,15){1.5}{12}
\ArrowLine(42,15)(72,15)
\ArrowLine(72,38)(23,38)
\CArc(23,29)(9,90,180)
\ArrowLine(14,29)(28,29)
\Gluon(0,29)(14,29){1.5}{4}
\GOval(36,22)(11,11)(0){0.9}
\Text(74,38)[l]{\scriptsize 4}
\Text(74,15)[l]{\scriptsize 3}
\end{picture}
}
& 10: $V_{g_1\overline{q}_4,p_0}^{\mathrm{ini,}S_1}$ \\
& 11: $V_{g_1\overline{q}_4,p_3}^{\mathrm{ini,}S_1}$ \\
& 12: $V_{g_1\overline{q}_4,p_2}^{\mathrm{ini,}S_1}$ \\
\\
\multirow{2}{*}{
\begin{picture}(77,44)(0,0)
\Line(36,30)(72,30)
\Line(36,28)(72,28)
\ArrowLine(0,29)(28,29)
\Gluon(0,15)(36,15){1.5}{12}
\ArrowLine(42,15)(72,15)
\Gluon(20,6)(72,6){1.5}{18}
\GlueArc(23,15)(9,180,270){-1.5}{4}
\GOval(36,22)(11,11)(0){0.9}
\Text(74,15)[l]{\scriptsize 3}
\Text(74,6)[l]{\scriptsize 4}
\end{picture}
}
& 13: $V_{g_2g_4,p_0}^{\mathrm{ini,}S_1}$ \\
& 14: $V_{g_2g_4,p_3}^{\mathrm{ini,}S_1}$ \\
& 15: $V_{g_2g_4,p_1}^{\mathrm{ini,}S_1}$ \\
\\
\multirow{2}{*}{
\begin{picture}(77,44)(0,0)
\Line(36,30)(72,30)
\Line(36,28)(72,28)
\ArrowLine(0,29)(28,29)
\Gluon(14,15)(36,15){1.5}{7}
\ArrowLine(42,15)(72,15)
\ArrowLine(0,15)(14,15)
\CArc(23,15)(9,180,270)
\ArrowLine(23,6)(72,6)
\GOval(36,22)(11,11)(0){0.9}
\Text(74,15)[l]{\scriptsize 3}
\Text(74,6)[l]{\scriptsize 4}
\end{picture}
}
& 16: $V_{q_2q_4,p_0}^{\mathrm{ini,}S_1}$ \\
& 17: $V_{q_2q_4,p_3}^{\mathrm{ini,}S_1}$ \\
& 18: $V_{q_2q_4,p_1}^{\mathrm{ini,}S_1}$ \\
\\
\multirow{2}{*}{
\begin{picture}(77,44)(0,0)
\Line(36,30)(72,30)
\Line(36,28)(72,28)
\ArrowLine(0,29)(28,29)
\Gluon(14,15)(36,15){1.5}{7}
\ArrowLine(42,15)(72,15)
\ArrowLine(0,15)(14,15)
\CArc(23,15)(9,180,270)
\ArrowLine(23,6)(72,6)
\GOval(36,22)(11,11)(0){0.9}
\Text(74,15)[l]{\scriptsize 4}
\Text(74,6)[l]{\scriptsize 3}
\end{picture}
}
& 19: $V_{q_2q_3,p_0}^{\mathrm{ini,}S_1}$ \\
& 20: $V_{q_2q_3,p_4}^{\mathrm{ini,}S_1}$ \\
& 21: $V_{q_2q_3,p_1}^{\mathrm{ini,}S_1}$ \\
\multicolumn{2}{c}{\small(equal quark flavors only)}
\end{tabular}
}
\end{minipage}
\hspace{\TODhspace}
\begin{minipage}[t]{\TODcolwidth}
\vspace{0pt}
\scalebox{\TODscalefactor}{
\begin{tabular}{cc}
\multicolumn{2}{c}{\bf $g+q\to c\overline{c}[n] + q:$} \\
\\
\multirow{2}{*}{
\begin{picture}(77,44)(0,0)
\Line(36,30)(72,30)
\Line(36,28)(72,28)
\ArrowLine(0,15)(28,15)
\Gluon(0,29)(36,29){1.5}{12}
\ArrowLine(42,15)(60,15)
\ArrowLine(60,15)(72,15)
\Gluon(56,15)(72,6){1.5}{6}
\GOval(36,22)(11,11)(0){0.9}
\Text(74,15)[l]{\scriptsize 3}
\Text(74,6)[l]{\scriptsize 4}
\end{picture}
}
& 1: $V_{q_3g_4,p_0}^{\mathrm{fin,}S_1}$ \\
& 2: $V_{q_3g_4,p_2}^{\mathrm{fin,}S_1}$ \\
& 3: $V_{q_3g_4,p_1}^{\mathrm{fin,}S_1}$ \\
\\
\multirow{2}{*}{
\begin{picture}(77,44)(0,0)
\Line(36,30)(72,30)
\Line(36,28)(72,28)
\ArrowLine(0,15)(28,15)
\Gluon(0,29)(36,29){1.5}{12}
\ArrowLine(42,15)(72,15)
\Gluon(57,29)(72,38){1.5}{6}
\GOval(36,22)(11,11)(0){0.9}
\Text(74,38)[l]{\scriptsize 4}
\Text(74,15)[l]{\scriptsize 3}
\end{picture}
}
& 4: $V_{p_0g_4,p_3}^{\mathrm{fin,}S_1}$ \\
& 5: $V_{p_0g_4,p_2}^{\mathrm{fin,}S_1}$ \\
& 6: $V_{p_0g_4,p_1}^{\mathrm{fin,}S_1}$ \\
\\
\multirow{2}{*}{
\begin{picture}(77,44)(0,0)
\Line(36,30)(72,30)
\Line(36,28)(72,28)
\Gluon(0,29)(36,29){1.5}{12}
\ArrowLine(0,15)(28,15)
\ArrowLine(42,15)(72,15)
\Gluon(20,38)(72,38){1.5}{18}
\GlueArc(23,29)(9,90,180){1.5}{4}
\GOval(36,22)(11,11)(0){0.9}
\Text(74,38)[l]{\scriptsize 4}
\Text(74,15)[l]{\scriptsize 3}
\end{picture}
}
& 7: $V_{g_1g_4,p_0}^{\mathrm{ini,}S_1}$ \\
& 8: $V_{g_1g_4,p_3}^{\mathrm{ini,}S_1}$ \\
& 9: $V_{g_1g_4,p_2}^{\mathrm{ini,}S_1}$ \\
\\
\multirow{2}{*}{
\begin{picture}(77,44)(0,0)
\Line(36,30)(72,30)
\Line(36,28)(72,28)
\Gluon(14,29)(36,29){1.5}{7}
\ArrowLine(0,15)(28,15)
\ArrowLine(42,15)(72,15)
\ArrowLine(23,38)(72,38)
\CArc(23,29)(9,90,180)
\ArrowLine(0,29)(14,29)
\GOval(36,22)(11,11)(0){0.9}
\Text(74,38)[l]{\scriptsize 4}
\Text(74,15)[l]{\scriptsize 3}
\end{picture}
}
& 10: $V_{q_1q_4,p_0}^{\mathrm{ini,}S_1}$ \\
& 11: $V_{q_1q_4,p_3}^{\mathrm{ini,}S_1}$ \\
& 12: $V_{q_1q_4,p_2}^{\mathrm{ini,}S_1}$ \\
\multicolumn{2}{c}{\small(equal quark flavors only)} \\
\\
\multirow{2}{*}{
\begin{picture}(77,44)(0,0)
\Line(36,30)(72,30)
\Line(36,28)(72,28)
\Gluon(14,29)(36,29){1.5}{7}
\ArrowLine(0,15)(28,15)
\ArrowLine(42,15)(72,15)
\ArrowLine(23,38)(72,38)
\CArc(23,29)(9,90,180)
\ArrowLine(0,29)(14,29)
\GOval(36,22)(11,11)(0){0.9}
\Text(74,38)[l]{\scriptsize 3}
\Text(74,15)[l]{\scriptsize 4}
\end{picture}
}
& 13: $V_{q_1q_3,p_0}^{\mathrm{ini,}S_1}$ \\
& 14: $V_{q_1q_3,p_4}^{\mathrm{ini,}S_1}$ \\
& 15: $V_{q_1q_3,p_2}^{\mathrm{ini,}S_1}$ \\
\\
\multirow{2}{*}{
\begin{picture}(77,44)(0,0)
\Line(36,30)(72,30)
\Line(36,28)(72,28)
\ArrowLine(0,15)(14,15)
\ArrowLine(14,15)(28,15)
\Gluon(0,29)(36,29){1.5}{12}
\ArrowLine(42,15)(72,15)
\Gluon(20,6)(72,6){1.5}{18}
\GlueArc(23,15)(9,180,270){-1.5}{4}
\GOval(36,22)(11,11)(0){0.9}
\Text(74,15)[l]{\scriptsize 3}
\Text(74,6)[l]{\scriptsize 4}
\end{picture}
}
& 16: $V_{q_2g_4,p_0}^{\mathrm{ini,}S_1}$ \\
& 17: $V_{q_2g_4,p_3}^{\mathrm{ini,}S_1}$ \\
& 18: $V_{q_2g_4,p_1}^{\mathrm{ini,}S_1}$ \\
\\
\multirow{2}{*}{
\begin{picture}(77,44)(0,0)
\Line(36,30)(72,30)
\Line(36,28)(72,28)
\Gluon(0,29)(36,29){1.5}{12}
\ArrowLine(14,15)(28,15)
\ArrowLine(42,15)(72,15)
\Gluon(0,15)(14,15){1.5}{4}
\CArc(23,15)(9,180,270)
\ArrowLine(72,6)(23,6)
\GOval(36,22)(11,11)(0){0.9}
\Text(74,15)[l]{\scriptsize 3}
\Text(74,6)[l]{\scriptsize 4}
\end{picture}
}
& 19: $V_{g_2\overline{q}_4,p_0}^{\mathrm{ini,}S_1}$ \\
& 20: $V_{g_2\overline{q}_4,p_3}^{\mathrm{ini,}S_1}$ \\
& 21: $V_{g_2\overline{q}_4,p_1}^{\mathrm{ini,}S_1}$ 
\end{tabular}
}
\end{minipage}
\hspace{\TODhspace}
\begin{minipage}[t]{\TODcolwidth}
\vspace{0pt}
\scalebox{\TODscalefactor}{
\begin{tabular}{c}
Additional dipoles for $P$-wave states\\
(include for all subprocesses): \\
\\
231: $V_{S_2,31}$ \\
232: $V_{S_2,32}$ \\
234: $V_{S_2,34}$ \\
\\
241: $V_{S_2,41}$ \\
242: $V_{S_2,42}$ \\
243: $V_{S_2,43}$ \\
\\
330: $V_{S_3,3}$ \\
340: $V_{S_3,4}$ \\
\end{tabular}
}
\end{minipage}
\end{center}
\caption{\label{fig:listofdipoles}
Numbered list of dipole terms for each of the occurring Born processes with $2\to 2$ kinematics. The diagrams related to the $V_{ij,k}^{\mathrm{ini,}S_1}$ and $V_{ij,k}^{\mathrm{fin,}S_1}$ terms indicate in which collinear or soft limits the latter contribute. Light-quark lines are to be summed over all quark flavors.}
\end{sidewaysfigure}

\subsection{Momentum mappings\label{sec:MomentumMapping}}

The subtraction term $d\hat\sigma_\mathrm{subtr}$ in Eq.~(\ref{eq:dipolesmoredetailedOne}) is defined in terms of $2\to 3$ kinematics variables, but the squared Born amplitudes contained therein describe $2\to 2$ processes. Therefore, we need to {\em map} the $2\to 2$ kinematics momenta $\{\tilde{p}_i\}$ of the squared Born matrix elements in Eq.~(\ref{eq:dipolesmoredetailed}) to the momenta $\{p_i\}$ of the $2\to 3$ kinematics processes. This means that we need relations of the kind
\begin{eqnarray}
 \tilde{p}_1=\tilde{p}_1(p_1,p_2,p_0,p_3,p_4) &\qquad& \tilde{p}_2=\tilde{p}_2(p_1,p_2,p_0,p_3,p_4), \nonumber \\
 \tilde{p}_0=\tilde{p}_0(p_1,p_2,p_0,p_3,p_4) &\qquad& \tilde{p}_3=\tilde{p}_3(p_1,p_2,p_0,p_3,p_4),
\end{eqnarray}
which fulfill certain conditions, at least $\tilde{p}_1^2=\tilde{p}_2^2=\tilde{p}_3^2=0$, $\tilde{p}_0^2=p_0^2$ and $\tilde{p}_1+\tilde{p}_2=\tilde{p}_0+\tilde{p_3}$. For dipoles that are to describe a limit where the outgoing momentum $p_3$ or $p_4$ is soft, or where $p_3$ and $p_4$ are collinear, we also need $\tilde{p}_1\to p_1$, $\tilde{p}_2\to p_2$, $\tilde{p}_0\to p_0$, and $\tilde{p}_3\to p_3 + p_4$ in those limits. For dipoles which are to describe an initial-state collinear limit where the final-state momentum $p_j$ is collinear to the initial-state momentum $p_a$, we need $\tilde{p}_a\to x p_a$, $\tilde{p}_b\to p_b$, $\tilde{p}_0\to p_0$ and $\tilde{p}_3\to p_3+p_4-(1-x)p_a$, where $x=(n\cdot p_a-n\cdot p_j)/(n\cdot p_a)$, $n$ is an arbitrary vector, and $p_b$ is the momentum of the incoming parton that is not splitting. Since we cannot fulfill all conditions at the same time, we need different mappings for different dipoles. The four kinds of momentum mappings we use in our study are the following.

For all dipoles that do not involve the quarkonium momentum $p_0$, we use the mapping that follows from Catani-Seymour chapters 5.2 and 5.3, and also 5.6 with $n=p_3+p_4$. With $p_a$ being an initial-state momentum, this mapping implies that
\begin{equation}
 \tilde{p}_a = x p_a, \qquad \tilde{p}_3 = p_3 + p_4 - (1-x) p_a, \qquad x = \frac{p_3\cdot p_a + p_4 \cdot p_a - p_3\cdot p_4}{p_3\cdot p_a + p_4 \cdot p_a}.
 \label{eq:MapCatSey}
\end{equation}
It satisfies the conditions for all the limits $p_3$ or $p_4$ soft, $p_3$ collinear to $p_4$, and $p_a$ collinear to either $p_3$ or $p_4$, and we refer to this mapping as {\em MapCS($p_a$)}.

For those dipoles that involve the quarkonium momentum $p_0$, an initial-state momentum $p_a$ and a massless final-state momentum $p_f$, we use the mapping
\begin{equation}
 \tilde{p}_a = x p_a, \qquad \tilde{p}_0 = p_0 + p_f - (1-x) p_a, \qquad x = \frac{p_f\cdot p_a + p_0\cdot p_a - p_0 \cdot p_f}{p_f\cdot p_a + p_0 \cdot p_a}
 \label{eq:MapWein6}
\end{equation}
of Phaf-Weinzierl chapters 6.1 and 6.2. It satisfies the conditions for the limits $p_f$ soft and $p_f$ collinear to $p_a$, and we refer to it as {\em MapPW6($p_a, p_f$)}.

If we have a dipole term involving the quarkonium momentum $p_0$ plus two final-state momenta $p_f$ and $p_g$, being $p_3$ and $p_4$ or vice versa, but we are only concerned with the limit $p_f$ soft, we use the mapping of Phaf-Weinzierl chapter 5.2, namely
\begin{equation}
 \tilde{p}_3 = \frac{1}{1-y} p_g, \qquad \tilde{p}_0 = p_0 + p_f - \frac{y}{1-y} p_g,
 \qquad y=\frac{p_0 \cdot p_f}{p_f\cdot p_g + p_0\cdot p_f + p_0 \cdot p_g},
 \label{eq:MapWein5.2}
\end{equation}
which we call {\em MapPW5.2($p_f$)}.

The case involving the final-state momenta $p_0$, $p_f$, and $p_g$, but with $p_0$ being the spectator, is more complicated, since here, in addition  to the condition for $p_f$ soft, also those for $p_g$ soft and for $p_f$ and $p_g$ collinear need to be fulfilled. The momentum mapping appropriate here is the one of Phaf-Weinzierl chapter 5.1 is
\begin{equation}
 \tilde{p}_3 = a p_f + b p_g + c p_0, \qquad \tilde{p}_0 = (1-a) p_f + (1-b) p_g + (1-c) p_0,
 \label{eq:MapWein5.1First}
\end{equation}
with
\begin{eqnarray}
 a&=& \frac{1}{N} \left( 1-u-\frac{c}{y \tilde{u}_0} \left(2y(1-u)-\tilde{u}_0((1-u)y^2+(1-u+u^2)y+u-2u^2)\right)\right),
 \nonumber \\
 b&=& \frac{1}{N} \left( u + \frac{c}{y \tilde{u}_0} \left(-2yu - \tilde{u}_0((1-2u-u^2)y+1-3u+2u^2)\right) \right), \nonumber \\
 c&=& \frac{y\tilde{u}_0}{(2u-1-y(1-u))^2\tilde{u}_0+4u(1-u)y}\left(2u(1-u)-\frac{N}{\sqrt{1-v}}\right),
 \nonumber \\
 N&=&u^2+(1-u)^2+(1-u)y,
 \label{eq:MapWein5.1Second}
\end{eqnarray}
where
\begin{equation}
 \tilde{u}_0= \frac{(p_f+p_g+p_0)^2-p_0^2}{(p_f+p_g+p_0)^2}, \qquad
 u = \frac{2p_f\cdot p_g + 2p_0\cdot p_g}{(p_f+p_g+p_0)^2-p_0^2}, \qquad
 v = \frac{p_f\cdot p_g ( p_0 \cdot p_f + \frac{p_0^2}{2} )}{p_f\cdot p_0 ( p_f\cdot p_g + p_0 \cdot p_g)}.
 \label{eq:MapWein5.1Third}
\end{equation}
We refer to this mapping as {\em MapPW5.1($p_f$)}.

We note that the dipole terms in the Catani-Seymour and Phaf-Weinzierl papers were constructed such that the spin correlation terms of the splitting gluons vanish when contracted with the splitting gluon's tilde momentum. This property is used in the analytic integrations as a simplification, but it assumes that the momentum mapping of the corresponding chapter is used. A momentum mapping alternative to Eq.~(\ref{eq:MapWein5.1First}) is given, for example, in Eq.~(5.9) of Ref.~\cite{Catani:2002hc}, which has the advantage of being symmetric in $p_f$ and $p_g$. But that mapping does not fulfill the contraction property of the dipole terms in Phaf-Weinzierl chapter 5.1, which we use.

\subsection{Phase space factorization\label{sec:phasspacefac}}

The phase space factorization $d\mathrm{PS}_3=d\mathrm{PS}_2 [dx] d\mathrm{PS}_{\mathrm{dipole}}$, with $d\mathrm{PS}_{\mathrm{dipole}}$ depending only on the external momenta involved in the respective dipole terms, is crucial to facilitate their analytic integrations over $d\mathrm{PS}_{\mathrm{dipole}}$. In the case of dipoles for final-state particles only, we have $d\mathrm{PS}_3=d\mathrm{PS}_2 d\mathrm{PS}_{\mathrm{dipole}}$, and, in the case of dipoles involving an initial-state parton with momentum $p_a$, the factorization is $d\mathrm{PS}_3=d\mathrm{PS}_2dx\,d\mathrm{PS}_{\mathrm{dipole}}$, where $x$ fulfills $\tilde{p}_a=xp_a$. The dipole factorization and the analytic integration can be found in the respective papers where the dipoles are given. The result of the analytic integration then only depends on the momenta $\{\tilde{p}_i\}$ and $x$. For the reader's convenience, we copy here the phase space parameterization of Phaf-Weinzierl chapters 5 and 6, only slightly adjusting the notation, since they will be the basis for our analytic integration of the $V_{S_2,ij}^{\beta}$ and $V_{S_3,j}^{\alpha\beta}$ terms in sections~\ref{sec:S2dipolesini}--\ref{sec:S3dipoles}.

The phase space parameterization used in Phaf-Weinzierl chapter 5, involving the quarkonium momentum $p_0$ and two final-state momenta $p_f$ and $p_g$ being $p_3$ and $p_4$ or vice versa, is
\begin{equation}
 d\mathrm{PS}_3(p_1+p_2\to p_0+p_3+p_4) =
 d\mathrm{PS}_2(\tilde{p}_1+\tilde{p}_2\to \tilde{p}_0+\tilde{p_3}) d\mathrm{PS}_\mathrm{dipole},
\end{equation}
with
\begin{equation}
 d\mathrm{PS}_{\mathrm{dipole}} = \frac{(4\pi)^{\epsilon-2}}{\Gamma(1-\epsilon)} \tilde{s}^{1-\epsilon} \tilde{u}_0^{2-2\epsilon} \int_0^1 du (1-u)^{1-2\epsilon} (1-\tilde{u}_0u)^{\epsilon-1} u^{1-2\epsilon} \int_0^1 dv v^{-\epsilon} (1-v)^{-\epsilon}, \label{eq:dPSdipoleWein5}
\end{equation}
where $\tilde{u}_0$, $u$, and $v$ are those of Eq.~(\ref{eq:MapWein5.1Third}) and $\tilde{s}=(\tilde{p}_0+\tilde{p}_3)^2$, which here equals $(p_0+p_3+p_4)^2$, so that $\tilde{u}_0=(\tilde{s}-4m_Q^2)/\tilde{s}$.

The phase space parameterization used in Phaf-Weinzierl chapter 6, involving the quarkonium momentum $p_0$, an initial-state momentum $p_a$ being $p_1$ or $p_2$ and a final-state momentum $p_f$ being $p_3$ or $p_4$, is
\begin{equation}
 d\mathrm{PS}_3(p_1+p_2\to p_0+p_3+p_4) =
 d\mathrm{PS}_2(\tilde{p}_1+\tilde{p}_2\to \tilde{p_0}+\tilde{p_3}) dx\,d\mathrm{PS}_\mathrm{dipole},
\end{equation}
with
\begin{equation}
 d\mathrm{PS}_{\mathrm{dipole}} = \frac{(4\pi)^{\epsilon-2}}{\Gamma(1-\epsilon)} (-\tilde{\psi}_a)^{1-\epsilon} x^{\epsilon-1} (1-x)^{1-2\epsilon} (1-\tilde{\chi_a}x)^{\epsilon-1} \int_0^1 dw w^{-\epsilon} (1-w)^{-\epsilon}, \label{eq:dPSdipoleWein6}
\end{equation}
where $x$ is that of Eq.~(\ref{eq:MapWein6}),
\begin{equation}
 w=\frac{p_a\cdot p_f (p_0\cdot p_f + \frac{p_0^2}{2})}{p_0\cdot p_f ( p_a\cdot p_f + p_0\cdot p_a)}, \qquad \tilde{\chi_a} = \frac{(p_0+p_f-p_a)^2}{(p_0+p_f-p_a)^2-p_0^2},
\end{equation}
and $\tilde{\psi}_a=(\tilde{p}_0-\tilde{p}_a)^2-4m_Q^2$, which here is equal to $(p_0+p_f-p_a)^2-p_0^2$, such that $\tilde{\chi_a}=(\tilde{\psi}_a+4m_Q^2)/\tilde{\psi}_a$.

\subsection{Integration of dipoles over dipole phase space}

\boldmath
\subsubsection{Integration of $V_{S_2,ij}$ terms: Initial-state case\label{sec:S2dipolesini}}
\unboldmath

To solve the dipole phase space integral of $V_{S_2,ij}^\beta$ given in Eq.~(\ref{eq:S2Vterm}) for an initial-state parton $i$, we use in the following the momentum mapping in Eq.~(\ref{eq:MapWein6}) and the parameterization of the dipole phase space in Eq.~(\ref{eq:dPSdipoleWein6}) with $p_a=p_i$ and $p_f=p_j$ in both equations. Since the integration result can only depend on the momenta $\tilde{p}_i$ and $\tilde{p}_0$, we start by decomposing
\begin{equation}
 \int d\mathrm{PS}_{\mathrm{dipole}} V_{S_2,ij}^{\beta\,(\mathrm{ini})} = C_1 \tilde{p}_i^\beta + C_2 \tilde{p}_0^\beta.
 \label{eq:IntS2IniFirst}
\end{equation}
Although the component proportional to $\tilde{p}_0^\beta$ will vanish upon contraction with $\epsilon_\beta(m_l)$ in Eq.~(\ref{eq:dipolesmoredetailed}), we still have to consider it here, since the integral itself does have this component. We determine $C_1$ by multiplying Eq.~(\ref{eq:IntS2IniFirst}) with $\tilde{p}_{i\beta}$ and $\tilde{p}_{0\beta}$ and solving the resulting system of linear equations and so obtain
\begin{eqnarray}
 C_1 &=& \frac{4g_s^2}{\tilde{p}_0\cdot\tilde{p}_i} \left[ -\frac{\tilde{p}_0\cdot p_i}{p_0\cdot p_j\;p_i\cdot p_j} + \frac{(p_0^2)^2 \;\tilde{p}_i\cdot p_j}{\tilde{p}_0\cdot \tilde{p}_i(p_0\cdot p_j)^3} -\frac{p_0^2 \;p_0\cdot p_i\;\tilde{p}_i\cdot p_j}{\tilde{p}_0\cdot \tilde{p}_i\; p_i\cdot p_j(p_0\cdot p_j)^2} \right. \nonumber \\
 && \left. -\frac{p_0^2 \;\tilde{p}_0\cdot p_j}{(p_0\cdot p_j)^3} + \frac{p_0\cdot p_i\;\tilde{p}_0\cdot p_j}{p_i\cdot p_j (p_0\cdot p_j)^2} \right].
\end{eqnarray}
Next, we apply the mapping in Eq.~(\ref{eq:MapWein6}), express all appearing scalar products in terms of $\tilde{\psi}_i$, $\tilde{\chi_i}$, $x$, $1-x$, $w$, and $1-\tilde{\chi_i}x$, and so obtain
\begin{equation}
C_1 = \frac{16g_s^2}{1-\tilde{\chi_i}x}\left[ \frac{wx}{\tilde{\psi}_i^2} - \frac{2x(1-\tilde{\chi_i}x)}{\tilde{\psi}_i^2(1-x)} + \frac{16m_Q^2x^2(1-\tilde{\chi_i}x)}{\tilde{\psi}_i^3(1-x)^2} - \frac{16m_Q^2wx^2}{\tilde{\psi}_i^3(1-x)}+\frac{64m_Q^4x^3w}{\tilde{\psi}_i^4(1-x)^2} \right].
\end{equation}
We now use the expression in Eq.~(\ref{eq:dPSdipoleWein6}) for the dipole phase space in Eq.~(\ref{eq:IntS2IniFirst}), perform the $w$ integration,
and expand the result in $\epsilon$ using
\begin{equation}
 2(1-x)^{-1-2\epsilon}=-\frac{1}{\epsilon}\delta(1-x)+\left(\frac{2}{1-x}\right)_+ + {\cal O}(\epsilon).
\end{equation}
The result through terms of order ${\cal O}(\epsilon^0)$ is then
\begin{eqnarray}
 \int d\mathrm{PS}_{\mathrm{dipole}} V_{S_2,ij}^{\beta\,(\mathrm{ini})} &=& \frac{g_s^2}{\pi^2\tilde{\psi}_i} \tilde{p}_i^\beta  \left[ \delta(1-x)\left(\frac{4\pi\mu_r^2}{m_Q^2}e^{-\gamma_E}\right)^\epsilon\left(-\frac{1}{\epsilon}-2-\ln\frac{4m_Q^2}{\tilde{\psi}_i^2}\right)  \right. \nonumber \\
 && \left.+ \frac{2x(1-\tilde{\chi_i})(2-x-\tilde{\chi_i}x)}{(1-\tilde{\chi_i}x)^2} \left(\frac{1}{1-x}\right)_+ + \frac{3(1-x)}{2(1-\tilde{\chi_i}x)^2} \right] + (\tilde{p}_0\,\mathrm{term}).\qquad\label{eq:dipsoft2iniresult}
\end{eqnarray}

\boldmath
\subsubsection{Integration of $V_{S_2,ij}$ terms: Final-state case\label{sec:S2dipolesfin}}
\unboldmath

To solve the dipole phase space integral of $V_{S_2,ij}^\beta$ given in Eq.~(\ref{eq:S2Vterm}) for a final-state parton $i$, we use in the following the momentum mapping in Eq.~(\ref{eq:MapWein5.2}) and the parameterization of the dipole phase space in Eq.~(\ref{eq:dPSdipoleWein5}) with $p_f=p_j$ and $p_g=p_i$ in both equations. The integration result can then only depend on the momenta $\tilde{p}_3$ and $\tilde{p}_0$, and we decompose
\begin{equation}
 \int d\mathrm{PS}_{\mathrm{dipole}} V_{S_2,ij}^{\beta\,(\mathrm{fin})} = C_3 \tilde{p}_3^\beta + C_4 \tilde{p}_0^\beta.
 \label{eq:IntS2FinFirst}
\end{equation}
Although the component proportional to $\tilde{p}_0^\beta$ will vanish upon contraction with $\epsilon_\beta(m_l)$ in Eq.~(\ref{eq:dipolesmoredetailed}), we still have to consider it here, since the integral itself does have this component. We determine $C_3$ by multiplying Eq.~(\ref{eq:IntS2FinFirst}) in turn with $\tilde{p}_{3\beta}$ and $\tilde{p}_{0\beta}$ and solving the resulting system of linear equations and so obtain
\begin{eqnarray}
 C_3 &=& \frac{8g_s^2}{\tilde{s}\tilde{u}_0} \left[ \frac{2(p_0^2)^2 \;\tilde{p}_3\cdot p_j}{\tilde{s}\tilde{u}_0 (p_0\cdot p_j)^3} - \frac{2p_0^2 \;\tilde{p}_3\cdot p_j\;p_0\cdot p_i}{\tilde{s}\tilde{u}_0 p_i\cdot p_j(p_0\cdot p_j)^2 } + \frac{2p_0^2 \;\tilde{p}_3\cdot p_i}{\tilde{s}\tilde{u}_0 p_0\cdot p_j\;p_i\cdot p_j} \right. \nonumber \\
 &&\left. - \frac{\tilde{p}_0\cdot p_i}{p_0\cdot p_j\;p_i\cdot p_j}-\frac{p_0^2 \;\tilde{p}_0\cdot p_j}{(p_0\cdot p_j)^3}  + \frac{p_0\cdot p_i\;\tilde{p}_0\cdot p_j}{p_i\cdot p_j(p_0\cdot p_j)^2 } \right].
\end{eqnarray}
Next, we apply the mapping in Eq.~(\ref{eq:MapWein5.2}), express all appearing scalar products in terms of $\tilde{s}$, $\tilde{u}_0$, $u$, $1-u$, $v$, and $1-\tilde{u}_0u$, and so obtain
\begin{eqnarray}
 C_3 &=& \frac{16 g_s^2v}{(1-\tilde{u}_0u)(1-u)^2\tilde{s}^2\tilde{u}_0} \left[ (1-u)^2+\frac{64m_Q^4}{\tilde{s}^2\tilde{u}_0^2} + \frac{16 m_Q^2(1-u)}{\tilde{s}\tilde{u}_0} - \frac{16 m_Q^2(1-\tilde{u}_0u)}{\tilde{s}\tilde{u}_0^2v} \right. \nonumber \\
 && \left. - \frac{2(1-\tilde{u}_0u)(1-u)}{\tilde{u}_0v}\right].
\end{eqnarray}
Using the expression in Eq.~(\ref{eq:dPSdipoleWein5}) for the dipole phase space in Eq.~(\ref{eq:IntS2FinFirst}), we can now do the integrations by identifying hypergeometric functions, which we then expand in $\epsilon$ using the program package {\tt HypExp} \cite{Huber:2007dx}. Our result through order ${\cal O}(\epsilon^0)$ is
\begin{eqnarray}
 \int d\mathrm{PS}_{\mathrm{dipole}} V_{S_2,ij}^{\beta\,(\mathrm{fin})} &=& \frac{g_s^2}{\pi^2\tilde{s}\tilde{u}_0}\left(\frac{4\pi\mu_r^2}{m_Q^2}e^{-\gamma_E}\right)^\epsilon \tilde{p}_3^\beta \left[\frac{1}{\epsilon}-\frac{1}{\tilde{u}_0}\ln\frac{4m_Q^2}{\tilde{s}}+3+\frac{1}{2}\ln\frac{64m_Q^{10}}{\tilde{s}^5\tilde{u}_0^4} \right] \nonumber \\
 && +(\tilde{p}_0\,\mathrm{term}).\label{eq:dipsoft2finresult}
\end{eqnarray}

\boldmath
\subsubsection{Integration of $V_{S_3,j}$ terms and incorporation of LDME renormalization counterterms\label{sec:S3dipoles}}
\unboldmath

To solve the dipole phase space integral of $V_{S_3,j}^{\alpha\beta}$, we again use the momentum mapping in Eq.~(\ref{eq:MapWein5.2}) and the parameterization of the dipole phase space in Eq.~(\ref{eq:dPSdipoleWein5}) with $p_f=p_j$ and $p_g=p_i$. Since the integration result can only depend on the momenta $\tilde{p}_3$ and $\tilde{p}_0$, we decompose
\begin{equation}
 \int d\mathrm{PS}_{\mathrm{dipole}} V_{S_3,j}^{\alpha\beta} = C_5 g^{\alpha\beta} + C_6 \tilde{p}_3^\alpha \tilde{p}_3^\beta + C_7 \tilde{p}_0^\alpha \tilde{p}_0^\beta + C_8(\tilde{p}_0^\alpha \tilde{p}_3^\beta + \tilde{p}_3^\alpha \tilde{p}_0^\beta).
 \label{eq:IntS3First}
\end{equation}
Although the components proportional to $\tilde{p}_0^\alpha$ and $\tilde{p}_0^\beta$ will vanish upon contraction with $\epsilon^\ast_\alpha(m_l)\epsilon_\beta(m_l)$ in Eq.~(\ref{eq:dipolesmoredetailed}), we still have to consider them here, since the integral itself does have these components. We determine $C_5$ and $C_6$ by multiplying Eq.~(\ref{eq:IntS3First}) with $g_{\alpha\beta}$, $\tilde{p}_{3\alpha}\tilde{p}_{3\beta}$, $\tilde{p}_{0\alpha}\tilde{p}_{0\beta}$, and $\tilde{p}_{0\alpha}\tilde{p}_{3\beta}$ and solving the resulting system of linear equations and so obtain
\begin{eqnarray}
 C_5 &=& \frac{4g_s^2}{(p_0\cdot p_j)^4} \left( -\frac{2(p_0^2)^2 (\tilde{p}_3\cdot p_j)^2}{(1-\epsilon)\tilde{s}^2\tilde{u}_0^2}  + \frac{2p_0^2\;\tilde{p}_0\cdot p_j\;\tilde{p}_3\cdot p_j}{(1-\epsilon)\tilde{s}\tilde{u}_0}- (p_0\cdot p_j)^2\right), \\
 C_6 &=& \frac{16g_s^2 p_0^2}{(p_0\cdot p_j)^4} \left(-\frac{(6-4\epsilon)(p_0^2)^2(\tilde{p}_3\cdot p_j)^2}{(1-\epsilon)\tilde{s}^4\tilde{u}_0^4} + \frac{(6-4\epsilon)p_0^2 \;\tilde{p}_0\cdot p_j\;\tilde{p}_3\cdot p_j}{(1-\epsilon)\tilde{s}^3\tilde{u}_0^3}  - \frac{(\tilde{p}_0\cdot p_j)^2}{\tilde{s}^2\tilde{u}_0^2} \right).
\end{eqnarray}
Next, we apply the mapping in Eq.~(\ref{eq:MapWein5.2}) and express all appearing scalar products in terms of $\tilde{s}$, $\tilde{u}_0$, $u$, $1-u$, $v$, and $1-\tilde{u}_0u$ and so obtain
\begin{eqnarray}
 C_5 &=& \frac{16g_s^2}{(1-\epsilon)(1-u)^2\tilde{u}_0^2}\left[ \frac{2p_0^2v}{(1-\tilde{u}_0u)\tilde{s}^3}-\frac{2p_0^2v^2}{(1-\tilde{u}_0u)^2}\left(\frac{p_0^2}{\tilde{s}^4}+\frac{(1-u)\tilde{u}_0}{\tilde{s}^3}\right) - \frac{1-\epsilon}{\tilde{s}^2}\right],
 \\
 C_6 &=&\frac{64 g_s^2 p_0^2 v^2}{(1-\tilde{u}_0u)^2(1-u)^2\tilde{s}^4\tilde{u}_0^2}\left[ \frac{(6-4\epsilon)p_0^2}{(1-\epsilon)\tilde{s}} \left( \frac{1-\tilde{u}_0u}{\tilde{u}_0^2v} - \frac{p_0^2}{\tilde{s}\tilde{u}_0^2} - \frac{1-u}{\tilde{u}_0} \right) \right. \nonumber \\
 && - \left. \frac{(1-\tilde{u}_0u)^2}{\tilde{u}_0^2v^2} + \frac{2(1-\tilde{u}_0u)(1-u)}{\tilde{u}_0v} -(1-u)^2 \right].
\end{eqnarray}
Using the expression in Eq.~(\ref{eq:dPSdipoleWein5}) for the dipole phase space in Eq.~(\ref{eq:IntS3First}), we can now do the integrations by identifying hypergeometric functions, which we then expand in $\epsilon$ using {\tt HypExp} \cite{Huber:2007dx}. Our result through order ${\cal O}(\epsilon^0)$ is
\begin{eqnarray}
 \int d\mathrm{PS}_{\mathrm{dipole}} V_{S_3,j}^{\alpha\beta} &=&  \frac{g_s^2}{12\pi^2m_Q^2} \left(\frac{4\pi\mu_r^2}{m_Q^2}e^{-\gamma_E}\right)^\epsilon g^{\alpha\beta} \left[ \frac{1}{\epsilon} + \frac{2}{3} - \frac{4m_Q^2}{\tilde{s}\tilde{u}_0}\ln\frac{16m_Q^6}{\tilde{s}^3\tilde{u}_0^2}-\frac{2\ln(2\tilde{u}_0)}{\tilde{u}_0}\right] \nonumber \\
 && + \frac{2g_s^2}{3\pi^2\tilde{s}^4\tilde{u}_0^3} \tilde{p}_3^\alpha \tilde{p}_3^\beta \left( 16m_Q^4 - \tilde{s}^2 - 8 m_Q^2\tilde{s} \ln\frac{4m_Q^2}{\tilde{s}}\right) + (\tilde{p}_0\,\mathrm{terms}).
 \label{eq:V3DipoleIntResult}
\end{eqnarray}

Let us now consider this result together with Eqs.~(\ref{eq:OpRenTransformed3PJOne})--(\ref{eq:OpRenTransformed1P1Two}) and (\ref{eq:dipolesmoredetailedOne}). For each partonic $2\to 3$ subprocess $a+b\to c\overline{c}[n]+X$, there is one (are two) contributions of $V_{S_3,j}^{\alpha\beta}$ if $X$ contains one (two) outgoing gluons and $n$ is a $P$-wave state. The divergence of each of these contributions equals $-\| |n,\mathrm{op.ren.}\rangle \|^2$ with the same partons $a$ and $b$. Noticing that $F_\mathrm{sym}(X)$ in the dipole subtraction term is 1 ($\frac{1}{2}$) if there is one (are two) outgoing gluon(s), but always 1 in the LDME renormalization contribution, we observe that the divergence in Eq.~(\ref{eq:V3DipoleIntResult}) is exactly canceled by the contributions from LDME renormalization. Thus, in our implementation, it is simplest to include the effects of the LDME renormalization by just using instead of Eq.~(\ref{eq:V3DipoleIntResult}) the expression
\begin{eqnarray}
 \left(\int d\mathrm{PS}_{\mathrm{dipole}} V_{S_3,j}^{\alpha\beta}\right)_\mathrm{+op.ren.} &=&  \frac{g_s^2}{12\pi^2m_Q^2} g^{\alpha\beta} \left( \frac{2}{3} - \frac{4m_Q^2}{\tilde{s}\tilde{u}_0}\ln\frac{16m_Q^6}{\tilde{s}^3\tilde{u}_0^2}-\frac{2\ln(2\tilde{u}_0)}{\tilde{u}_0}-\ln\frac{m_Q^2}{\mu_\Lambda^2}\right) \nonumber \\
 && + \frac{2g_s^2}{3\pi^2\tilde{s}^4\tilde{u}_0^3} \tilde{p}_3^\alpha \tilde{p}_3^\beta \left( 16m_Q^4 - \tilde{s}^2 - 8 m_Q^2\tilde{s} \ln\frac{4m_Q^2}{\tilde{s}}\right),\label{eq:dipsoft3oprenresult}
\end{eqnarray}
which is then finite.

\boldmath
\subsubsection{Integration of $V_{ij,k}^{S_1,\mathrm{ini}}$ and $V_{ij,k}^{S_1,\mathrm{fin}}$ terms and incorporation of mass factorization counterterm}
\unboldmath

There is one subtlety related to the dipole terms of $V_{ij,k}^{S_1,\mathrm{ini}}$ in the initial-state collinear limits $\tilde{p}_i\to xp_i$. In the second bracket of Eq.~(\ref{eq:dipolesubgeneral}), there is then an apparent mismatch because $d\hat{\sigma}_\mathrm{subtr}$ involves parton $i$ with momentum $p_i$, while $d\hat\sigma_\mathrm{virtual}$ and $d\hat\sigma_\mathrm{MFC}$ involve initial-state parton $(ij)$ with momentum $xp_i$ instead. Thus, special care has to be exercised regarding the differing color and polarization averaging and flux factors. In order to facilitate the singularity cancellation, it is, therefore, convenient to rewrite the contribution of the $V_{ij,k}^{S_1,\mathrm{ini}}$ terms in $d\hat\sigma_\mathrm{subtr}$ when appearing in the second bracket of Eq.~(\ref{eq:dipolesubgeneral}) as
\begin{eqnarray}
&&d\hat{\sigma}_\mathrm{subtr}(a+b\to Q\overline{Q}[n]+X; V_{ij,k}^{S_1,\mathrm{ini}}	) = - d\mathrm{PS}_2 dx \frac{1}{N_\mathrm{col}(n) N_\mathrm{pol}(n)}\,\frac{1}{2x(p_1+p_2)^2} \nonumber \\
&&\quad\times \frac{F_\mathrm{sym}(X) n_\mathrm{col}(i) n_\mathrm{pol}(i)}{n_\mathrm{col}(a) n_\mathrm{pol}(a) n_\mathrm{col}(b) n_\mathrm{pol}(b) n_\mathrm{col}((ij)) n_\mathrm{pol}((ij))}
   \langle n, \mathrm{Born} | {\cal V}^{\mathrm{ini,}S_1}_{ij,k} \frac{\mathbf{T}_{(ij)} \mathbf{T}_k}{\mathbf{T}_{(ij)}^2} | n, \mathrm{Born} \rangle,\quad\label{eq:ininew}
\end{eqnarray}
with the terms
\begin{equation}
 {\cal V}^{\mathrm{ini,}S_1}_{ij,k} = \int d\mathrm{PS}_\mathrm{dipole} \frac{n_\mathrm{pol}((ij))}{n_\mathrm{pol}(i)}\frac{1}{2p_i\cdot p_j} V^{\mathrm{ini,}S_1}_{ij,k},\label{eq:intini}
\end{equation}
analytically calculated in the Catani-Seymour and Phaf-Weinzierl papers.

Now we consider Eq.~(\ref{eq:ininew}) together with Eq.~(\ref{eq:MFC}). Using again the trick $\sum_{\substack{k=0\\k\neq i,j}}^4 \frac{\mathbf{T_{(ij)}\mathbf{T}_k}}{\mathbf{T}_{(ij)}^2}=-1$ and noticing that the effect of double contributions due to $j=3,4$ is balanced by the symmetry factor $F_\mathrm{sym}(X)=\frac{1}{2}$ for two gluons in the final state, we observe that we can incorporate the effect of the mass factorization counterterm completely by using the expressions
\begin{equation}
 \left( {\cal V}^{\mathrm{ini,}S_1}_{ij,k}\right)_{+\mathrm{MFC}} = {\cal V}^{\mathrm{ini,}S_1}_{ij,k} +  \frac{g_s^2}{8\pi^2} \left( \frac{4\pi\mu_r^2}{\mu_f^2}e^{-\gamma_E}\right)^\epsilon \frac{1}{\epsilon} P_{i,(ij)}^+(x), \label{eq:intinimfc}
\end{equation}
instead of Eq.~(\ref{eq:intini}). For the reader's convenience, we collect the expressions for $\left( {\cal V}^{\mathrm{ini,}S_1}_{ij,k}\right)_{+\mathrm{MFC}}$ and those for
\begin{equation}
 {\cal V}^{\mathrm{fin,}S_1}_{ij,k} = \int d\mathrm{PS}_\mathrm{dipole} \frac{1}{2p_i\cdot p_j} V^{\mathrm{fin,}S_1}_{ij,k}\label{eq:intfin}
\end{equation}
in Appendix~\ref{sec:AppIntCSPW}.

\section{Implementation and numerical tests\label{sec:Implementation}}

\subsection{Implementation of phase space cuts}

The master formula (\ref{eq:dipolesubgeneral}) describes a total cross section. The observables we aim to calculate are, however, cross sections with specific kinematic cuts, for example, on the transverse momentum $p_T$ or the rapidity $y$ of the quarkonium. To this end, we define
\begin{equation}
 \tilde{p}_T^2 = \frac{(4m_Q^2-\tilde{t})(\tilde{s}+\tilde{t})}{\tilde{s}}-4m_Q^2,
 \qquad
 \tilde{y} = \ln \frac{\tilde{s} + \tilde{t}}{\tilde{x}_2 \sqrt{S\left(\tilde{p}_T^2+4m_Q^2\right)}},
\end{equation}
with
\begin{equation}
 \tilde{s}=(\tilde{p}_1+\tilde{p}_2)^2, \qquad \tilde{t}=(\tilde{p}_0-\tilde{p}_1)^2, \qquad S=(p_A+p_B)^2, \qquad \tilde{p}_2=\tilde{x}_2 p_B,
\end{equation}
where $p_A$ and $p_B$ are the momenta of the incoming hadrons.
For all momentum mappings, we then have $\tilde{p}_T\to p_T$ and $\tilde{y}\to y$ in all singular limits. We can thus refine Eq.~(\ref{eq:dipolesubgeneral}) to include the kinematic constraints. For example, the cross section with a phase space cut $p_T>p_{T,\mathrm{min}}$ is calculated as
\begin{eqnarray}
 \int d\hat{\sigma} &=& \int d\mathrm{PS}_3 \left[ \frac{d\hat{\sigma}_\mathrm{real}}{d\mathrm{PS}_3} \theta(p_T-p_{T,\mathrm{min}})-\frac{d\hat{\sigma}_\mathrm{subtr}}{d\mathrm{PS}_3} \theta(\tilde{p}_T-p_{T,\mathrm{min}}) \right]
 \nonumber \\
 &&+\int d\mathrm{PS}_2 \left[ \frac{d\hat{\sigma}_\mathrm{virtual}+d\hat{\sigma}_\mathrm{MFC}+d\hat{\sigma}_\mathrm{op.\,ren.}}{d\mathrm{PS}_2} \theta(p_T-p_{T,\mathrm{min}}) \right. \nonumber \\
 && \left.+ [dx] \theta(\tilde{p}_T-p_{T,\mathrm{min}}) \int d\mathrm{PS}_\mathrm{dipole} \frac{d\hat{\sigma}_\mathrm{subtr}}{d\mathrm{PS}_3} \right].
 \label{eq:dipolesubwithcuts}
\end{eqnarray}
In the first line of Eq.~(\ref{eq:dipolesubwithcuts}), we integrate over the complete three-particle phase space and implement the $\theta$ functions explicitly. The $\theta$ functions then cut out different regions of the three-particle phase space, depending on the momentum mappings used in each dipole term. This worsens the convergence of the numerical Monte-Carlo integration, but the $\theta$ functions coincide close to all singular regions, so that the cancellations of the divergent terms take place. We note that the strong-coupling constant in our implementation is usually evaluated at a renormalization scale that is chosen to depend on kinematic variables of the produced quarkonium, e.g., $\alpha_s(p_T^2)$. We then have to substitute $\alpha_s(\tilde{p}_T^2)$ in $d\hat{\sigma}_\mathrm{subtr}$. As for the contributions in the third line of Eq.~(\ref{eq:dipolesubwithcuts}), the analytic integration of the subtraction term over the dipole phase space $d\mathrm{PS}_\mathrm{dipole}$ is not affected by the additionally imposed phase space cuts, since $\tilde{p}_T$ only depends on the momenta $\{\tilde{p}_i\}$.

Equation~(\ref{eq:dipolesubwithcuts}) allows for the evaluation of binned cross section distributions, e.g., in $p_T$ and/or $y$, which can be directly compared with experimental data. Refining the binning of such histograms yields approximations to smooth cross section distributions. To evaluate the latter exactly, however, one needs to replace the $\theta$ functions in Eq.~(\ref{eq:dipolesubwithcuts}) by $\delta$ functions, which renders the implementation of the cancellation of divergences quite nontrivial. We leave the elaboration of this for future work.

\subsection{Numerical tests}

\begin{table}
\begin{center}
\begin{varwidth}{\linewidth}
\begin{scriptsize}
\begin{verbatim}
State: 3S18          real corr.    dipoles                               real corr.    dipoles

ggg 1,2,3 coll:      2.0041E+08   2.0041E+08        gdd 1,2,3 coll:      1.0428E+06   1.0423E+06
ggg 4,5,6 coll:      1.7072E+06   1.7071E+06        gdd 7,8,9 coll:      2.1892E+06   2.1888E+06
ggg 13,14,15 coll:   1.6866E+11   1.6866E+11        gdd 10,11,12 coll:   2.1572E+05   2.1561E+05
ggg 16,17,18 coll:   1.6879E+07   1.6883E+07        gdd 13,14,15 coll:   2.1516E+09   2.1516E+09
ggg 19,20,21 coll:   1.6429E+10   1.6429E+10        gdd 16,17,18 coll:   3.1073E+09   3.1073E+09
ggg 22,23,24 coll:   1.6866E+11   1.6866E+11        gdd 19,20,21 coll:   6.3923E+08   6.3922E+08
ggg 25,26,27 coll:   1.6866E+11   1.6866E+11        gdd p4 soft:         1.4382E+11   1.4382E+11
ggg 28,29,30 coll:   1.6429E+10   1.6429E+10        dDg 1,2,3 coll:      4.3257E+06   4.3257E+06
ggg p3 soft:         1.2357E+12   1.2357E+12        dDg 4,5,6 coll:      4.1623E+04   4.1624E+04
ggg p4 soft:         1.2357E+12   1.2357E+12        dDg 13,14,15 coll:   8.5966E+08   8.5966E+08
dgd 1,2,3 coll:      2.9102E+05   2.9110E+05        dDg 16,17,18 coll:   8.5894E+04   8.5885E+04
dgd 7,8,9 coll:      3.1100E+05   3.1092E+05        dDg 19,20,21 coll:   1.7684E+08   1.7684E+08
dgd 10,11,12 coll:   6.3894E+04   6.3923E+04        dDg 22,23,24 coll:   8.5966E+08   8.5966E+08
dgd 13,14,15 coll:   2.1869E+10   2.1869E+10        dDg 25,26,27 coll:   8.5966E+08   8.5966E+08
dgd 16,17,18 coll:   2.1516E+09   2.1516E+09        dDg 28,29,30 coll:   1.7684E+08   1.7684E+08
dgd 19,20,21 coll:   2.1516E+09   2.1516E+09        dDg p3 soft:         9.3760E+09   9.3759E+09
dgd p4 soft:         1.6118E+10   1.6119E+10        dDg p4 soft:         9.3760E+09   9.3759E+09
\end{verbatim}
\end{scriptsize}
\end{varwidth}
\end{center}
\caption{\label{tab:dipoletest3s18}Numerical test of the dipole terms for the partonic subprocesses $gg\to c\overline{c}[{^3S}_1^{[8]}] + gg$ ({\tt ggg}), $dg\to c\overline{c}[{^3S}_1^{[8]}] + dg$ ({\tt dgd}), $gd\to c\overline{c}[{^3S}_1^{[8]}] + dg$ ({\tt gdd}), and $d\overline{d}\to c\overline{c}[{^3S}_1^{[8]}] + gg$ ({\tt dDg}). The coding is as in Fig.~\ref{fig:listofdipoles}.}
\end{table}

\begin{table}
\begin{center}
\begin{varwidth}{\linewidth}
\begin{scriptsize}
\begin{verbatim}
State: 3P21            real corr.    dipoles         soft S1      soft S2      soft S3

gg2cCgg p3 soft:       4.2128E+10   4.2130E+10      3.7994E+10  -1.5081E+09   5.6437E+09
gg2cCgg p4 soft:       4.2128E+10   4.2130E+10      3.7994E+10  -1.5081E+09   5.6437E+09
dg2cCdg p4 soft:       1.8035E+08   1.8039E+08      4.1168E+07  -2.2504E+07   1.6173E+08
gd2cCdg p4 soft:       5.5345E+09   5.5347E+09      4.9728E+09  -3.0023E+08   8.6211E+08
dD2cCgg p3 soft:       8.9238E+07   8.9252E+07      1.6245E+07   8.2038E+06   6.4803E+07
dD2cCgg p4 soft:       8.9234E+07   8.9252E+07      1.6245E+07   8.2038E+06   6.4803E+07


State: 1P18            real corr.    dipoles         soft S1      soft S2      soft S3

gg2cCgg p3 soft:       1.1062E+11   1.1062E+11      1.1374E+11  -1.1212E+10   8.0929E+09
gg2cCgg p4 soft:       1.1062E+11   1.1062E+11      1.1374E+11  -1.1212E+10   8.0929E+09
dg2cCdg p4 soft:       3.6100E+08   3.6101E+08      3.5567E+08  -2.1288E+07   2.6633E+07
gd2cCdg p4 soft:       1.4423E+10   1.4424E+10      1.4588E+10  -1.2053E+09   1.0405E+09
dD2cCgg p3 soft:       1.1020E+08   1.1020E+08      1.0914E+08  -6.5828E+06   7.6456E+06
dD2cCgg p4 soft:       1.1018E+08   1.1020E+08      1.0914E+08  -6.5828E+06   7.6456E+06
\end{verbatim}
\end{scriptsize}
\end{varwidth}
\end{center}
\caption{\label{tab:dipoletestp}Numerical test of the dipole terms for the partonic subprocesses $gg\to c\overline{c}[{^3P}_2^{[1]}; {^1P}_1^{[8]}] + gg$ ({\tt gg2cCgg}), $dg\to c\overline{c}[{^3P}_2^{[1]}; {^1P}_1^{[8]}] + dg$ ({\tt dg2cCdg}), $gd\to c\overline{c}[{^3P}_2^{[1]}; {^1P}_1^{[8]}] + dg$ ({\tt gd2cCdg}), and $d\overline{d}\to c\overline{c}[{^3P}_2^{[1]}; ^1P_1^{[8]}] + gg$ ({\tt dD2cCgg}) in the limits where $p_3$ and $p_4$ are soft. The contributions of the dipoles involving $V^{S_1}$, $V^{S_2}$, and $V^{S_3}$ are shown separately.}
\end{table}

We now numerically verify the implementation of the individual unintegrated dipole terms. The subtraction term must match all the real-correction squared matrix elements in the respective limits. Three dipoles are always needed to reproduce a collinear limit, many dipoles to reproduce a soft limit. As an illustration, we generate certain phase space points close to the singularities and evaluate there both the real-correction squared matrix elements and the corresponding dipole terms. Our results are presented in Tables~\ref{tab:dipoletest3s18} and \ref{tab:dipoletestp}. From there we observe that the squared matrix elements of the real corrections are indeed nicely matched by the corresponding subtraction terms constructed as described above for all the partonic subprocesses, Fock states, and kinematic limits considered.

To obtain meaningful numerical checks of the implementation of the integrated dipole terms, also beyond self-consistency, it is indispensable to compare with results obtained using phase space slicing. This is even more so the case for checks of the implementation of dipole subtraction in calculations of physical observables of quarkonium production. Extensive such tests have successfully been performed, for all our integrated dipole terms and several phenomenological applications. Presenting them in detail would require to explain the anatomy of the implementation of phase space slicing in NLO NRQCD calculations, which reaches beyond the scope of this paper. We will report on such comparisons in a separate communication \cite{bk}, in which we will also quantitatively describe how dipole subtraction outperforms phase space slicing with respect to numerical precision and computing time.

\section{Summary\label{sec:Summary}}

We devised an implementation of a subtraction scheme appropriate for studies of inclusive quarkonium production at NLO in the NRQCD factorization approach, based on the dipole subtraction scheme of Refs.~\cite{Catani:1996vz,Phaf:2001gc}. We needed to take special care of the specific structure of the bound-state amplitudes and to include additional subtraction terms in the case of $P$-wave states. Our implementation passes all intrinsic tests and yields results consistent with our previous phase space slicing implementation, which it outruns both in terms of accuracy and speed.

\section*{Acknowledgments}

This work was supported in part by the German Federal Ministry for Education and Research BMBF through Grant No.\ 05H15GUCC1 and by the German Research Foundation DFG through Grant No.\ KN~365/12-1.

\begin{appendix}

\section{Summary of integrated Catani-Seymour and Phaf-Weinzierl dipoles\label{sec:AppIntCSPW}}

In this appendix, we collect the expressions through order ${\cal O}(\epsilon^0)$ for the integrated Catani-Seymour and Phaf-Weinzierl dipoles that we need in our study. The mass factorization counterterms are directly included here according to our definitions in Eqs.~(\ref{eq:intinimfc}) and (\ref{eq:intfin}). $g_i$, $q_i$, and $\overline{q}_i$ stand for a gluon, light quark, and antiquark with momentum $p_i$, and we further introduce $\tilde{\xi}_i=(\tilde{p}_3-\tilde{p}_i)^2$. Note that our expressions for ${\cal V}_{ij,p_{1\,\mathrm{or}\,2}}^{\mathrm{ini,}S_1}$ imply that $n=p_3+p_4$ as in Catani-Seymour chapter 5.6, in line with Table~\ref{tab:DipoleOverview}. The expressions for the integrated $V_{S_2,ij}$ and $V_{S_3,j}$ terms, including the operator renormalization counterterms in the latter case, can be found in Eqs.~(\ref{eq:dipsoft2iniresult}), (\ref{eq:dipsoft2finresult}), and (\ref{eq:dipsoft3oprenresult}). We have

\begin{eqnarray}
&&\left( {\cal V}_{g_i\overline{q}_j,p_0}^{\mathrm{ini,}S_1}\right)_\mathrm{+MFC} = \frac{g_s^2}{8\pi^2} \frac{1}{2}  \left[ 2x(1-x)-(x^2+(1-x)^2)\left(\ln\frac{x(1-\tilde{\chi_i}x)}{(1-x)^2} + \ln \frac{\mu_f^2}{-\tilde{\psi}_i}\right) \right],\\
&&\left( {\cal V}_{q_iq_j,p_0}^{\mathrm{ini,}S_1}\right)_\mathrm{+MFC} = \frac{g_s^2}{8\pi^2} C_F  \left[ x-\left(x+2\frac{1-x}{x}\right)\left(\ln\frac{x(1-\tilde{\chi_i}x)}{(1-x)^2}   + \ln \frac{\mu_f^2}{-\tilde{\psi}_i}\right) \right],\\
&&\left({\cal V}_{q_ig_j,p_0}^{\mathrm{ini,}S_1}\right)_\mathrm{+MFC} = \frac{g_s^2}{8\pi^2} C_F  \left(\frac{4\pi\mu_r^2}{-\tilde{\psi}_i}e^{-\gamma_E}\right)^\epsilon  \left\{ -\left(\frac{2}{1-x}\right)_+ \ln\frac{\mu_f^2}{-\tilde{\psi}_i} + 4 \left(\frac{\ln(1-x)}{1-x}\right)_+ \right. \nonumber \\
 &&\quad{} - 2(\ln x + \ln(2-\tilde{\chi_i}x)) \left(\frac{1}{1-x}\right)_+ + 1-x + (1+x)\left(\ln\frac{x(1-\tilde{\chi_i}x)}{(1-x)^2}   + \ln \frac{\mu_f^2}{-\tilde{\psi}_i}\right)
\nonumber \\
 &&\quad{} + \delta(1-x) \left[ \frac{1}{\epsilon^2} + \frac{1}{\epsilon}\left(\ln(2-\tilde{\chi_i})  + \frac{3}{2} \right) +\frac{\pi^2}{12} + 2\ln(1-\tilde{\chi_i})\ln(2-\tilde{\chi_i}) + 2\Li(\tilde{\chi_i}-1)\right. \nonumber \\
  &&\quad{}-\left.\left. \frac{1}{2}\ln^2(2-\tilde{\chi_i}) + \frac{3}{2}\ln\frac{-\tilde{\psi}_i}{\mu_f^2} \right] \right\},\\
&&\left({\cal V}_{g_ig_j,p_0}^{\mathrm{ini,}S_1}\right)_\mathrm{+MFC} = \frac{g_s^2}{8\pi^2} 2 C_A  \left(\frac{4\pi\mu_r^2}{-\tilde{\psi}_i}e^{-\gamma_E}\right)^\epsilon  \left\{ \left(-\ln x - \ln(2-\tilde{\chi_i}x)-\ln\frac{\mu_f^2}{-\tilde{\psi}_i}\right) \left(\frac{1}{1-x}\right)_+  \right. \nonumber \\
 &&\quad{} + 2\left(\frac{\ln(1-x)}{1-x}\right)_+ + \left(2-\frac{1}{x}-x+x^2\right)\left(\ln \frac{x(1-\tilde{\chi_i}x)}{(1-x)^2} + \ln \frac{\mu_f^2}{-\tilde{\psi}_i}\right)  \nonumber \\
 &&\quad{} + \delta(1-x) \left[ \frac{1}{2\epsilon^2} + \frac{1}{\epsilon}\left(\frac{1}{2}\ln(2-\tilde{\chi_i}) + \frac{11}{12} - \frac{n_f}{6C_A} \right) + \ln\frac{-\tilde{\psi}_i}{\mu_f^2} \left( \frac{11}{12}-\frac{n_f}{6C_A}\right)  \right. \nonumber \\
 &&\quad{}+ \left.\left.  \Li(\tilde{\chi_i}-1)+\ln(1-\tilde{\chi_i})\ln(2-\tilde{\chi_i})  - \frac{1}{4}\ln^2(2-\tilde{\chi_i}) + \frac{\pi^2}{24}  \right] \right\},\\
&&\left({\cal V}_{q_iq_j,p_{1\,\mathrm{or}\,2}}^{\mathrm{ini,}S_1}\right)_\mathrm{+MFC} = \frac{g_s^2}{8\pi^2} C_F \left[\frac{1+(1-x)^2}{x} \ln\frac{(x-1)\tilde{\xi}_i}{x\mu_f^2} + x\right],\\
&&\left({\cal V}_{g_iq_j,p_{1\,\mathrm{or}\,2}}^{\mathrm{ini,}S_1}\right)_\mathrm{+MFC} = \frac{g_s^2}{8\pi^2} \frac{1}{2} \left[\left(x^2+(1-x)^2\right) \ln\frac{(x-1)\tilde{\xi}_i}{x\mu_f^2} + 2x(1-x)\right],\\
&&\left({\cal V}_{q_ig_j,p_{1\,\mathrm{or}\,2}}^{\mathrm{ini,}S_1}\right)_\mathrm{+MFC} = \frac{g_s^2}{8\pi^2} C_F \left(\frac{4\pi\mu_r^2}{\tilde{s}}e^{-\gamma_E}\right)^\epsilon \left\{ -\left(\frac{1+x^2}{1-x}\right)_+ \ln\frac{x\mu_f^2}{\tilde{s}} + \left(\frac{4\ln(1-x)}{1-x}\right)_+  \right. \nonumber \\
 &&\quad{}+ \left.1-x- (1-x)\ln\frac{(x-1)\tilde{\xi}_i}{\tilde{s}} + \delta(1-x) \left[ \frac{1}{\epsilon^2} + \frac{3}{2\epsilon} +2\Li\left(\frac{\tilde{\chi_i}\tilde{\psi}_i}{\tilde{s}+\tilde{\psi}_i}\right)+\frac{\pi^2}{12} \right]\right\},\\
&&\left({\cal V}_{g_ig_j,p_{1\,\mathrm{or}\,2}}^{\mathrm{ini,}S_1}\right)_\mathrm{+MFC} = \frac{g_s^2}{8\pi^2} 2 C_A \left(\frac{4\pi\mu_r^2}{\tilde{s}}e^{-\gamma_E}\right)^\epsilon \left\{ \left(\frac{1-x}{x}-1+x(1-x)\right)\ln\frac{(x-1)\tilde{\xi}_i}{x\mu_f^2}\right. \nonumber \\
 &&\quad{} -\left(\frac{1}{1-x}\right)_+ \ln \frac{x\mu_f^2}{\tilde{s}} + \left(\frac{2\ln(1-x)}{1-x}\right)_+ + \delta(1-x)\left[ \frac{1}{2 \epsilon^2} + \frac{1}{\epsilon}\left(\frac{11}{12}-\frac{n_f}{6C_A}\right) \right. \nonumber \\
 &&\quad{}+ \left. \left. \Li\left(\frac{\tilde{\chi_i}\tilde{\psi}_i}{\tilde{s}+\tilde{\psi}_i}\right)+\frac{\pi^2}{24} +  \left(\frac{11}{12}-\frac{n_f}{6C_A}\right)\ln\frac{\tilde{s}}{\mu_f^2} \right] \right\},\\
&&\left({\cal V}_{q_ig_j,p_{3\,\mathrm{or}\,4}}^{\mathrm{ini,}S_1}\right)_\mathrm{+MFC} = \frac{g_s^2}{8\pi^2}C_F \left(\frac{4\pi\mu_r^2}{-\tilde{\xi}_i}e^{-\gamma_E}\right)^\epsilon \left\{ -\left(\frac{1+x^2}{1-x}\right)_+ \ln \frac{\mu_f^2x}{-\tilde{\xi}_i} + \left( \frac{4\ln(1-x)}{1-x}\right)_+ \right. \nonumber \\
 &&\quad{}- \left.  \frac{2\ln(2-x)}{1-x} -(1+x)\ln(1-x)+1-x + \delta(1-x) \left[ \frac{1}{\epsilon^2} + \frac{3}{2\epsilon} +\frac{\pi^2}{12} \right] \right\},\\
&&\left({\cal V}_{g_i\overline{q}_j,p_{3\,\mathrm{or}\,4}}^{\mathrm{ini,}S_1}\right)_\mathrm{+MFC} = \frac{g_s^2}{8\pi^2}\frac{1}{2} \left[ (x^2+(1-x)^2)\left(\ln(1-x)-\ln\frac{\mu_f^2x}{-\tilde{\xi}_i}\right) + 2x(1-x)\right],\\
&&\left({\cal V}_{q_iq_j,p_{3\,\mathrm{or}\,4}}^{\mathrm{ini,}S_1}\right)_\mathrm{+MFC} = \frac{g_s^2}{8\pi^2} C_F \left[ \frac{1+(1-x)^2}{x} \left(\ln(1-x)-\ln\frac{\mu_f^2x}{-\tilde{\xi}_i}\right) + x \right],\\
&&\left({\cal V}_{g_ig_j,p_{3\,\mathrm{or}\,4}}^{\mathrm{ini,}S_1}\right)_\mathrm{+MFC} = \frac{g_s^2}{8\pi^2} 2 C_A \left(\frac{4\pi\mu_r^2}{-\tilde{\xi}_i}e^{-\gamma_E}\right)^\epsilon \left\{ -\left(\frac{1}{1-x}\right)_+ \ln\frac{\mu_f^2x}{-\tilde{\xi}_i} + \left(\frac{2\ln(1-x)}{1-x}\right)_+  \right. \nonumber \\
&&\quad{} - \frac{\ln(2-x)}{1-x} +\left(-1+x(1-x)+\frac{1-x}{x}\right)\left(\ln(1-x)-\ln\frac{\mu_f^2x}{-\tilde{\xi}_i}\right)  \nonumber \\
&&\quad{}+ \left. \delta(1-x) \left[ \frac{1}{2\epsilon^2}  + \frac{1}{\epsilon}\left(\frac{11}{12}-\frac{n_f}{6C_A}\right) +\frac{\pi^2}{24} +\left(\frac{11}{12}-\frac{n_f}{6C_A}\right)\ln\frac{-\tilde{\xi}_i}{\mu_f^2} \right] \right\},\\
&&{\cal V}_{p_0g_j,p_{k=1\,\mathrm{or}\,2}}^{\mathrm{fin,}S_1} = \frac{g_s^2}{8\pi^2} C_F  \left(\frac{4\pi\mu_r^2}{-\tilde{\psi}_k}e^{-\gamma_E}\right)^\epsilon  \left\{ \left(\frac{2}{1-x}\right)_+ \left(\ln\frac{2-\tilde{\chi_k}x}{1-\tilde{\chi_k}x} - \frac{(1-\tilde{\chi_k})x^2}{1-\tilde{\chi_k}x}\right) \right. \nonumber \\
 &&\quad{} + \delta(1-x) \left[ \frac{1}{\epsilon}\left( 1+\ln\frac{1-\tilde{\chi_k}}{2-\tilde{\chi_k}} \right) +2-\frac{\pi^2}{3} + \ln(1-\tilde{\chi_k})+ \frac{1}{2}\ln^2(1-\tilde{\chi_k})\right. \nonumber \\
 &&\quad{}+ \left.\left.   \frac{1}{2}\ln^2(2-\tilde{\chi_k})-2\ln(1-\tilde{\chi_k})\ln(2-\tilde{\chi_k})  - 2\Li(\tilde{\chi_k}-1) \right] \right\},\\
&&{\cal V}_{p_0g_j,p_{3\,\mathrm{or}\,4}}^{\mathrm{fin,}S_1} = \frac{g_s^2}{8\pi^2} C_F  \left(\frac{4\pi\mu_r^2}{\tilde{s}\tilde{u}_0^2}e^{-\gamma_E}\right)^\epsilon \left[ \frac{1}{\epsilon}(1+\ln(1-\tilde{u}_0)) +4+\ln(1-\tilde{u}_0)-4\Li(\tilde{u}_0) \right. \nonumber \\
 &&\quad{}- \left. \frac{1}{2}\ln^2(1-\tilde{u}_0) \right],\\
 &&{\cal V}_{q_i\overline{q}_j,p_{k=1\,\mathrm{or}\,2}}^{\mathrm{fin,}S_1} = \frac{g_s^2}{8\pi^2} \frac{1}{2} \left(\frac{4\pi\mu_r^2}{-\tilde{\xi}_k}e^{-\gamma_E}\right)^\epsilon \left\{ \frac{2}{3}\left(\frac{1}{1-x}\right)_+ + \delta(1-x)\left[ -\frac{2}{3\epsilon}-\frac{10}{9} \right]\right\},\\
&&{\cal V}_{q_ig_j,p_{k=1\,\mathrm{or}\,2}}^{\mathrm{fin,}S_1} = \frac{g_s^2}{8\pi^2} C_F \left(\frac{4\pi\mu_r^2}{-\tilde{\xi}_k}e^{-\gamma_E}\right)^\epsilon \left\{ -\left(\frac{2\ln(1-x)}{1-x}\right)_+-\frac{3}{2}\left(\frac{1}{1-x}\right)_+ + \frac{2\ln(2-x)}{1-x}\right. \nonumber \\
 &&\quad{}+ \left.  \delta(1-x)\left[\frac{1}{\epsilon^2}+\frac{3}{2\epsilon} + \frac{7}{2}- \frac{7}{12}\pi^2 \right]\right\},\\
&&{\cal V}_{g_ig_j,p_{k=1\,\mathrm{or}\,2}}^{\mathrm{fin,}S_1} = \frac{g_s^2}{8\pi^2}2C_A \left(\frac{4\pi\mu_r^2}{-\tilde{\xi}_k}e^{-\gamma_E}\right)^\epsilon \left\{ -\left(\frac{2\ln(1-x)}{1-x}\right)_+-\frac{11}{6}\left(\frac{1}{1-x}\right)_+ + \frac{2\ln(2-x)}{1-x} \right. \nonumber \\
 &&\quad{} + \left. \delta(1-x) \left[ \frac{1}{\epsilon^2} + \frac{11}{6\epsilon} + \frac{67}{18} - \frac{7}{12}\pi^2 \right]\right\},\\
&&{\cal V}_{g_ig_j,p_0}^{\mathrm{fin,}S_1} = \frac{g_s^2}{8\pi^2} 2 C_A  \left(\frac{4\pi\mu_r^2}{\tilde{s}\tilde{u}_0^2}e^{-\gamma_E}\right)^\epsilon \left[ \frac{1}{\epsilon^2} + \frac{11}{6\epsilon} - \frac{5}{6}\pi^2 + \frac{67}{12} - \frac{1-\tilde{u}_0}{3\tilde{u}_0^2} + 2 \Li(\tilde{u}_0) \right. \nonumber \\
  &&\quad{}- \left.  \frac{1}{6}\left( \frac{1-\tilde{u}_0}{\tilde{u}_0^3} (2-\tilde{u}_0) + 11 \frac{1-\tilde{u}_0}{\tilde{u}_0}\right) \ln (1-\tilde{u}_0) - \frac{\pi^2}{12} \right],\\
&&{\cal V}_{q_i\overline{q}_j,p_0}^{\mathrm{fin,}S_1} = \frac{g_s^2}{8\pi^2} \frac{1}{2} \left(\frac{4\pi\mu_r^2}{\tilde{s}\tilde{u}_0^2}e^{-\gamma_E}\right)^\epsilon  \left[ - \frac{2}{3\epsilon} -\frac{11}{6} +\frac{2(1-\tilde{u}_0)}{3\tilde{u}_0^2} \right. \nonumber \\
 &&\quad{}+ \left.  \frac{1-\tilde{u}_0}{3\tilde{u}_0^3}(2\tilde{u}_0^2-\tilde{u}_0+2)\ln(1-\tilde{u}_0) \right],\\
&&{\cal V}_{q_ig_j,p_0}^{\mathrm{fin,}S_1} = \frac{g_s^2}{8\pi^2} C_F  \left(\frac{4\pi\mu_r^2}{\tilde{s}\tilde{u}_0^2}e^{-\gamma_E}\right)^\epsilon  \left[ \frac{1}{\epsilon^2} + \frac{3}{2\epsilon} + \frac{19}{4} - \frac{11}{12}\pi^2 + \frac{1}{2\tilde{u}_0} + 2\Li(\tilde{u}_0)\right. \nonumber \\
 &&\quad{}+ \left.  \frac{(1-\tilde{u}_0)(1-3\tilde{u}_0)}{2\tilde{u}_0^2}\ln(1-\tilde{u}_0)  \right].
\end{eqnarray}

\end{appendix}

\end{document}